\documentclass[
 superscriptaddress,
 reprint,
 amsmath,amssymb,
 aps,
 prd
]{revtex4-1}

\usepackage{graphicx} 
\usepackage{dcolumn}
\usepackage{bm}
\usepackage{color}
\usepackage{hyperref}
\usepackage[mathlines]{lineno}

\begin{document}

\preprint{APS/123-QED}

\title{Probing gluon TMD with reconstructed and tagged heavy flavor hadron pairs at the EIC} 

\author{Xin Dong}
\affiliation{Lawrence Berkeley National Laboratory, Berkeley, CA 94720, USA}

\author{Yuanjing Ji}
\affiliation{Lawrence Berkeley National Laboratory, Berkeley, CA 94720, USA}

\author{Matthew Kelsey}
\affiliation{Wayne State University, Detroit, MI 48202, USA}
\affiliation{Lawrence Berkeley National Laboratory, Berkeley, CA 94720, USA}

\author{Sooraj Radhakrishnan}
\email{skradhakrishnan@lbl.gov}
\affiliation{Kent State University, Kent, OH 44242, USA}
\affiliation{Lawrence Berkeley National Laboratory, Berkeley, CA 94720, USA}

\author{Ernst Sichtermann}
\affiliation{Lawrence Berkeley National Laboratory, Berkeley, CA 94720, USA}

\author{Yuxiang Zhao}
\affiliation{Institute of Modern Physics, Chinese Academy of Sciences, Lanzhou, Gansu Province 730000, China}
\affiliation{University of Chinese Academy of Sciences, Beijing 100049, China}
\affiliation{Key Laboratory of Quark and Lepton Physics (MOE) and Institute of Particle Physics, Central China Normal University, Wuhan 430079, China}

\date{\today}

\begin{abstract}
   Study of the transverse structure of the proton is one of the major physics goals of the upcoming Electron Ion Collider (EIC). The gluon transverse momentum dependent distributions (TMD) form an essential focus of this effort and are important towards understanding the angular momentum contribution to proton spin as well as QCD factorization. However, very limited experimental constraints on the gluon TMD exist currently. As the heavy quark production in lepton-nucleon DIS gets a dominant contribution from the photon-gluon-fusion process, heavy quark production makes an attractive tool to probe gluon distributions in nucleons. In this paper we present a study of heavy flavor hadron pair reconstruction at a future EIC detector with MAPS based inner tracking and vertexing subsystems to constrain gluon TMD. We utilize the excellent track pointing resolution provided by the detector to exclusively reconstruct heavy flavor hadron pairs via their hadronic decay channels and also to develop a heavy flavor hadron tagging algorithm. Statistical uncertainty projections on azimuthal asymmetries corresponding to gluon TMD at the EIC is evaluated. The heavy flavor tagging is found to substantially enhance the purity of heavy flavor hadron pair selection, and the statistical precision of the measurement compared to that from exclusive reconstruction. The correlation between the azimuthal angle of the transverse momentum of the gluon initiating the process and that of the corresponding heavy flavor hadron pair was also studied and found to be well correlated. This study opens up heavy flavor hadron pair measurements as an attractive channel to access gluon TMD at the EIC. 

\end{abstract}


\maketitle

\section{Introduction}
The quarks and gluons inside hadrons have a non-trivial distribution in transverse momentum. In polarized or unpolarized hadrons they can be spin polarized with the direction and magnitude of the polarization depending on their transverse momentum, flavor and in case of polarized hadrons, the hadron polarization. These spin-orbit couplings provide for a rich transverse structure of the hadrons and also challenge our fundamental understanding of Quantum Chromodynamics (QCD) in many ways~\cite{GrossePerdekamp:2015xdx,Sivers:1989cc,Boer:2010zf,Collins:2011zzd}. Experimentally, they are known to contribute to the transverse single spin asymmetries (SSAs) measured in inclusive hadron production in hadron-hadron collisions and SIDIS processes~\cite{E581:1991eys,Krueger:1998hz,Allgower:2002qi,STAR:2003lxu,PHENIX:2005jxc,STAR:2008ixi,BRAHMS:2008doi,STAR:2012hth,STAR:2012ljf,AnDY:2013att,PHENIX:2013wle,PHENIX:2014qwb,HERMES:2013quo,HERMES:2004mhh,JeffersonLabHallA:2011ayy,JeffersonLabHallA:2013mjr,COMPASS:2010hbb}. The upcoming Electron Ion Collider (EIC), has the study of these correlations and the transverse structure of the proton as one of the major scientific goals~\cite{Accardi:2012qut,AbdulKhalek:2021gbh}. 

The transverse structure of the proton can be studied using Transverse Momentum Dependent parton distribution functions (TMD PDFs), assuming TMD factorization~\cite{GrossePerdekamp:2015xdx,Barone:2010zz,Collins:2011zzd}. However, unlike the collinear PDFs, the TMD can be process dependent from the initial and final state processes required to preserve color gauge invariance, thus breaking universality~\cite{Brodsky:2002cx,Buffing:2013kca,Collins:2002kn}. TMD factorization also needs to be studied and established for different processes. TMD are thus fundamentally interesting quantities to study in non-perturbative QCD. One of the most well known and studied spin-related TMD is the Sivers TMD~\cite{Sivers:1989cc,DAlesio:2015fwo}. The Sivers TMD quantifies the left-right asymmetry in the distribution of partons, with respect to the plane formed by the momentum and spin directions of the proton, in a transversely polarized proton ('Sivers effect'). The Sivers effect is considered to be important in understanding the parton angular momentum and thus potentially the angular momentum contribution to nucleon spin~\cite{COMPASS:2017ezz,Sivers:1989cc,Sivers:2006rg,Bacchetta:2011gx}. Measurements and experimental constraints on TMD are mostly limited to quarks~\cite{Anselmino:2005ea,Anselmino:2008sga,Barone:2010gk,Echevarria:2020hpy,Bury:2021sue,Bacchetta:2020gko,Cammarota:2020qcw,Gamberg:2022kdb}. Gluon TMD on the other hand are poorly constrained from experimental data currently~\cite{Boer:2015vso,DAlesio:2015fwo,DAlesio:2018rnv,COMPASS:2017ezz}. 


Some estimates for gluon Sivers TMD from both hadron-hadron collision and SIDIS data exist. Fits to SIDIS data on transverse asymmetry in pion and kaon production from HERMES~\cite{HERMES:2009lmz} and COMPASS~\cite{COMPASS:2005csq} experiments using quark Sivers TMD were found to nearly satisfy the Burkardt's sum rule, which requires that the total transverse momentum of all partons in a transversely polarized nucleon vanishes~\cite{Anselmino:2008sga,Brodsky:2006ha,Burkardt:2004ur}. This leaves little room for a gluon Sivers effect, but allows for gluon Sivers TMD $\sim ~1/N_{\mathrm{c}}$ ($N_{\mathrm{c}}$ = 3 being the number of colors) times the valance quark Sivers TMD within uncertainties~\cite{Boer:2015vso}. The SIDIS measurements considered are at small negative squared momentum transfer ($Q^2$) values and moderately large momentum fraction $x$ (0.01 $< x <$ 0.4), and therefore does not allow to draw conclusions on the gluon Sivers TMD at large $Q^2$ and small $x$. Estimates of the gluon Sivers effect has also been made using transverse SSA in $\pi^0$ production and inclusive pion and $D$ meson production from recent PHENIX measurements~\cite{PHENIX:2013wle,PHENIX:2017wbv} in polarized $pp$ collisions. In the large $x$ region, $0.05 < x < 0.3$, they estimate the normalized (to the unpolarized gluon TMD) gluon Sivers effect to be small, a few percent of the positivity bound~\cite{DAlesio:2015fwo,DAlesio:2018rnv}. Positivity bound is a trivial theoretical bound satisfied by polarized TMD~\cite{Boer:2015vso,Mulders:2000sh}. The uncertainties allow gluon Sivers TMD of the order of $\sim ~1/N_{\mathrm{c}}$ times the valance quark Sivers TMD, as with the SIDIS case. The evaluation also carries the caveat that for inclusive processes in proton proton collisions TMD factorization has not been proven~\cite{DAlesio:2015fwo}. On the other hand, from the more recent COMPASS measurement in DIS with transversely polarized protons and deuterons, the extracted transverse asymmetry $A_{\mathrm{UT}}^{\mathrm{PGF}}$ from the photon gluon fusion (PGF) process, which directly probes the gluon distributions, gives a large value of $-0.23$ at large $x$ ($x \sim ~0.1$)~\cite{COMPASS:2017ezz}. The measurement is more than 2 standard deviations away from zero and supports the possibility of a sizable gluon Sivers effect.

The most promising process to study gluon Sivers effect in electron-proton scattering is the heavy flavor pair-production process, $ep^{\uparrow} \rightarrow e'c\bar{c}X$~\cite{Boer:2015vso}. Selecting on the heavy (charm or bottom) quarks allows to tag the PGF process and minimize contributions from other subprocesses. Unlike di-jet or di-hadron production which receive sizeable contributions from the quark channel, particularly at large $x$ where the quark contribution dominates, the heavy quark pair production is dominated by gluon channel at all $x$ and gets only minor contribution from the quark channel~\cite{Zheng:2018ssm}. Also, TMD factorization might be easier to prove for the process and is proven to hold for SIDIS processes where the hadron transverse momentum is much less than $Q^{2}$~\cite{Ji:2004wu}. The EIC would allow probing the process over a large range in the $x-Q^2$ space. The observation of a transverse SSA in $ep^{\uparrow} \rightarrow e'c\bar{c}X$ process would be a smoking gun for gluon Sivers effect at the EIC~\cite{Boer:2015vso}. 

Experimental constraints on gluon TMD other than the Sivers TMD and the unpolarized gluon TMD are practically non-existent. However, recent theoretical studies on gluon polarization in unpolarized $ep$ collisions, where the spin-orbit coupling can give rise to a linear polarization of the gluons, showed sizable values for the transverse anisotropy observables associated with the linearly polarized gluon TMD~\cite{Boer:2010zf,Boer:2016fqd}. Here also, the process $ep \rightarrow e'c\bar{c}X$ is presented as providing the ideal opportunity to study the linearly polarized gluon TMD. The magnitude of the transverse anisotropy associated with the linearly polarized gluon TMD for heavy flavor pairs was predicted to be of the order of 10\% in the kinematic regions accessible at the EIC. These, and the fact that $ep \rightarrow e'c\bar{c}X$ provides the clean way to tag the PGF process and thus to probe the gluon distributions, makes the heavy flavor hadron pair measurements of particular interest at the EIC. In addition to studying the gluon distributions, heavy flavor hadron pair measurements can also be an attractive tool to study modifications to parton fragmentation from nuclear matter effects in electron - ion collisions, similar to its utility in heavy-ion collisions~\cite{Vogt:2018oje,Citron:2018lsq}. 

The gluon Sivers TMD can be studied using transverse single spin asymmetry ($A_{\mathrm{UT}}$) measurements in polarized $ep$ collisions. $A_{\mathrm{\mathrm{UT}}}(x,Q^2)$ is defined in the standard way as~\cite{Zheng:2018ssm}, 
\begin{equation}
   A_{\mathrm{\mathrm{UT}}}(x,Q^2) = \frac{\sigma_{\mathrm{L}}(x,Q^2) - \sigma_{\mathrm{R}}(x,Q^2)}{\sigma_{\mathrm{L}}(x,Q^2) + \sigma_{\mathrm{R}}(x,Q^2)},
   \label{eq1}
\end{equation}
where $\sigma_{\mathrm{L(R)}}$ are the cross sections for particle-of-interest production with spin polarized in the direction opposite to (same as) the spin of the proton, and $x$ is the momentum fraction of the parton. The asymmetry is directly related to the gluon Sivers effect, $A_{\mathrm{\mathrm{UT}}}(x,Q^2) \propto f_{\mathrm{1T}}^{\mathrm{\perp g}}(x,Q^2)/f_{\mathrm{1}}^{\mathrm{g}}(x,Q^2)$, where $f_{\mathrm{1T}}^{\mathrm{\perp g}}$ and $f_{\mathrm{1}}^{\mathrm{g}}$ are the gluon Sivers TMD and the unpolarized gluon TMD, respectively. The TMD of linearly polarized gluons can be probed through the measurements of the azimuthal anisotropy of the produced heavy flavor hadron pair momentum, $\langle \cos{(2\phi_{\mathrm{T}})} \rangle$, where $\phi_{\mathrm{T}}$ is the azimuthal angle corresponding to the summed momenta of the two heavy flavor hadrons in the pair. The asymmetry is related to the linearly polarized TMD as~\cite{Boer:2016fqd}, 
\begin{equation}
   \lvert \langle \cos(2\phi_{\mathrm{T}}) \rangle \rvert_{x,Q^2,k_{\mathrm{T}}} \propto \frac{q_{\mathrm{T}}^2}{2M_{\mathrm{p}}^2}\frac{h_{\mathrm{1}}^{\mathrm{\perp,g}}(x,Q^2,k_{\mathrm{T}})}{f_\mathrm{1}^{\mathrm{g}}(x,Q^2,k_{\mathrm{T}})},
\end{equation}
where $q_{\mathrm{T}}$ is the sum of the momenta of the heavy flavor hadrons in the pair, $M_{\mathrm{p}}$ is the proton mass and $k_{\mathrm{T}}$ is the gluon transverse momentum. 

In this study we present a detailed simulation study, using a realistic detector performance for an EIC detector, of using heavy flavor hadron pairs to study gluon TMD at the EIC. Previous simulation studies have looked at heavy flavor hadron pair measurements at the EIC~\cite{Zheng:2018ssm}. However, these were without including detector effects and for the heavy flavor channel, limited to explicit reconstruction of heavy flavor hadrons. We use a detector design and detector resolution parameters corresponding to a silicon tracker with vertexing and inner tracking layers using MAPS sensors~\cite{Arrington:2021yeb}. We also take advantage of the excellent track pointing resolution provided by the detector to develop a heavy flavor tagging algorithm to tag heavy flavor hadrons through their displaced decay tracks. The tagging algorithm is found to have good efficiency and purity and significantly improves the precision the heavy flavor hadron pair $A_{\mathrm{UT}}$ measurements. The tagging algorithm developed is hadron blind utilizing only the track pointing capabilities. The paper is organized as follows. The simulation setup and detector specifications and performance are discussed in section~\ref{sec:sim}. The details of heavy flavor hadron pair reconstruction and tagging studies are presented in section~\ref{sec:reco} and detailed statistical uncertainty projections for the gluon TMD observables are provided in section~\ref{sec:proj}.   

\section{Simulation setup}
\label{sec:sim}
Heavy flavor hadrons have very short decay lengths, $\sim$100$\mu$m. The major factor that enhances the performance of the detector in explicit reconstruction or tagging of the heavy flavor hadron decays is the track pointing resolution provide by the detector, quantified usually by the variance of the Distance of Closest Approach (DCA) of tracks to the vertex. Monolithic Active Pixel Sensors (MAPS) based tracking and vertexing systems have been employed with great success in attaining excellent track pointing resolutions and reconstruction performance for heavy flavor hadrons~\cite{Contin:2017mck,Fantoni:2020iyr}. We have studied a detector design with an all silicon MAPS based tracking system for an EIC detector~\cite{Arrington:2021yeb}. It consists of a barrel detector with 3 $\times$ 2 layers of MAPS based silicon pixels, covering $|\eta| <$ 1, and five MAPS based silicon pixel planes each in the forward and backward regions covering approximately 1 $< |\eta| <$ 3. The detector proposal from the ECCE collaboration~\cite{Ecce:2021ec} (and also  the ATHENA proposal~\cite{Athena:2021at}) has a very similar tracking system as the all silicon design for the vertexing layers and inner tracking layers and disks, with some variations on the technology choice for the outer tracking layers and number of disks on forward and backward directions. 

The performance of charm hadron reconstruction and utilizing heavy flavor hadron pairs for studying asymmetries corresponding to TMD are done using a fast simulation setup. The particle level momentum and vertex position from the event generator are smeared using parameterized single track momentum and pointing resolutions corresponding to those generated using full GEANT4~\cite{GEANT4:2002zbu} simulations of the all silicon detector. Such a procedure is chosen to save computing power, as the heavy flavor measurements are statistics hungry. The procedure has been used for impact projection studies on gluon nuclear PDFs and gluon helicity distributions using heavy flavor hadrons with the all silicon tracker at the EIC~\cite{Kelsey:2021gpk,Anderle:2021hpa}. A closure test was performed to validate the fast simulation procedure, comparing charm hadron reconstruction performance using the fast simulation with single track momentum and track pointing resolution parameters from GEANT4 simulation of the all-silicon tracker and the full reconstruction of the charm hadrons directly from the GEANT4 simulation of tracker. Good agreement has been found between the overall reconstruction efficiency of charm hadrons and also for the distribution of the different topological variables characterizing the charm hadron decay~\cite{Kelsey:2021gpk}. The detailed single particle momentum and track pointing resolutions as well as particle identification (PID) capabilities used in the fast-simulation study are provided in Table.1 of the article~\cite{Kelsey:2021gpk}. The momentum and track pointing resolutions and particle identification (PID) capabilities utilized are similar to those in the EIC Yellow Report~\cite{AbdulKhalek:2021gbh}. The primary vertex (PV) resolution is determined by the GEANT4 simulation of the tracker and is evaluated as a function of event multiplicity, and then utilized for smearing the PV in the fast simulation.

Events are generated for $ep$ collisions using PYTHIA 6.4 event generator~\cite{Sjostrand:2006za} that describes well the charm cross-section measurements at HERA, and was used for previous simulation studies at the EIC~\cite{Zheng:2018ssm,Kelsey:2021gpk}. The simulation studies are done with electron beams with energy 18 GeV and proton beams with 275 GeV. The kinematic variables used are defined in the conventional way. In the one-photon-exchange approximation, the incoming electron of four momentum $e$ emits a virtual photon of momentum $q = e - e'$, with $e'$ being the four momentum of the outgoing electron. The virtual photon then interacts with the hadron beam with four momentum $p$. The hadron momentum is taken to be along the positive $z$ direction in the simulation. The Bjorken scaling variable is $x_{\mathrm{B}} = Q^2/(2p.q)$ and $Q^2 \equiv -q^2$ is negative of the square of the four momentum transfer. The inelasticity $y = p · q/(p · e)$. The events generated are required to have $Q^2 >$ 1 GeV$^2$ and 0.005 $< y <$ 0.95. For the purposes of these studies we do not include any radiative corrections to the incoming/scattered lepton. The details of the event generation set up are same as those used in the previous study~\cite{Kelsey:2021gpk}. The coordinates in the study are kept in lab frame.

\begin{figure}[h]
\center{\includegraphics[width=1.0\columnwidth]{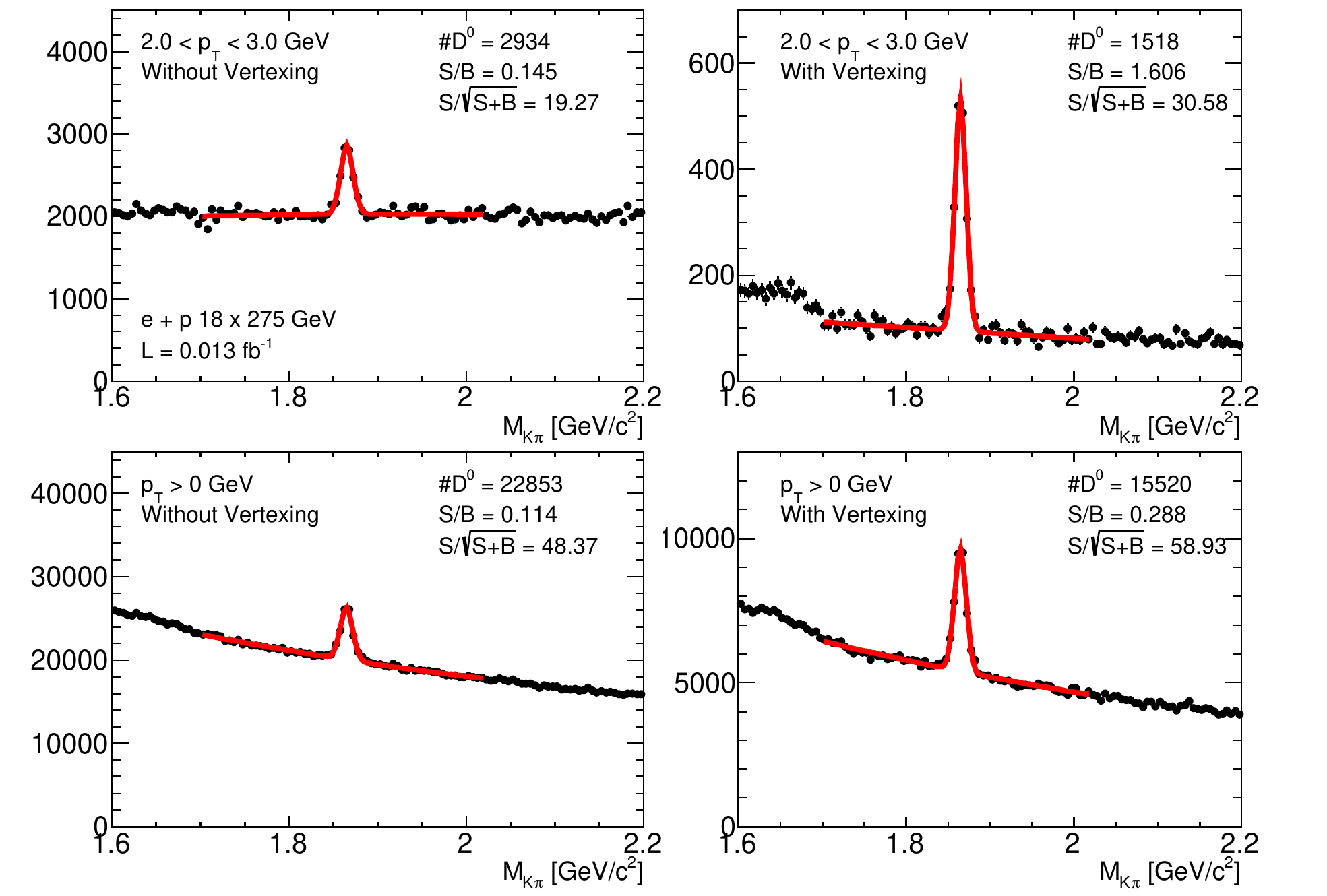}}
   \caption{The $D^{0}$ invariant mass distributions in two $p_{\mathrm{T}}$ bins without (left) and with (right) cuts on topological variable distributions characterizing the $D^0$ decay vertex.
}
\label{fig:d0mass}
\end{figure}

An example of $D^0$ meson reconstruction using the detector simulation is shown in Fig~\ref{fig:d0mass}. $D^0$ mesons are reconstructed through the $D^0 \rightarrow K^- \pi^+$ channel and its charge conjugate. The $D^0$ mesons have a $c\tau$ of approximately 120 $\mu$m~\cite{Zyla:2020zbs}. The excellent track pointing resolution offered by the tracking detector allows to place selection cuts on variables characterizing the decay topology to improve the signal to background ratio and the signal significance. The track pointing resolution utilized in the fast-simulation is better in the transverse ($r-\phi$) plane than the $z$-direction, and for these studies the topological variables used for selection cuts are in the transverse plane. Future studies could be conducted with more mature detector design parametrizations, combining the selections in both $r-\phi$ and $z$ directions to improve the performance. Cuts on the transverse distance between the PV and the reconstructed vertex of the $K\pi$ pair (decaylength), the distance of closest approach (DCA) between the $K\pi$ pairs (pairDCA) and the cosine of the angle between the $D^0$ candidate momentum and the vector joining the PV and $D^0$ candidate vertex in the transverse plane ($\cos{\theta}$) are utilized. Fig~\ref{fig:d0mass} shows the $K\pi$ pair invariant mass distribution without any cuts on the decay topology and with the cuts. The background is from random $K\pi$ pair combinations from tracks in the event which include all stable charged particle tracks from the PYTHIA event, within the detector acceptance. The cuts employed for $D^0$ reconstruction are decaylength $>$ 40 $\mu$m, pairDCA $<$ 150 $\mu$m, $\cos{\theta}$ $>$ 0.98. The topological selection cuts improves the S/B ratio and the signal significance considerably, particularly for $D^0$ higher transverse momentum $p_{\mathrm{T}}$. This provides a data sample with higher signal significance and reduces systematic uncertainties associated with the signal extraction. 

\section{Heavy flavor pair reconstruction and tagging}
\label{sec:reco}
\begin{figure}[htbp]
\center{
\includegraphics[width=0.9\columnwidth]{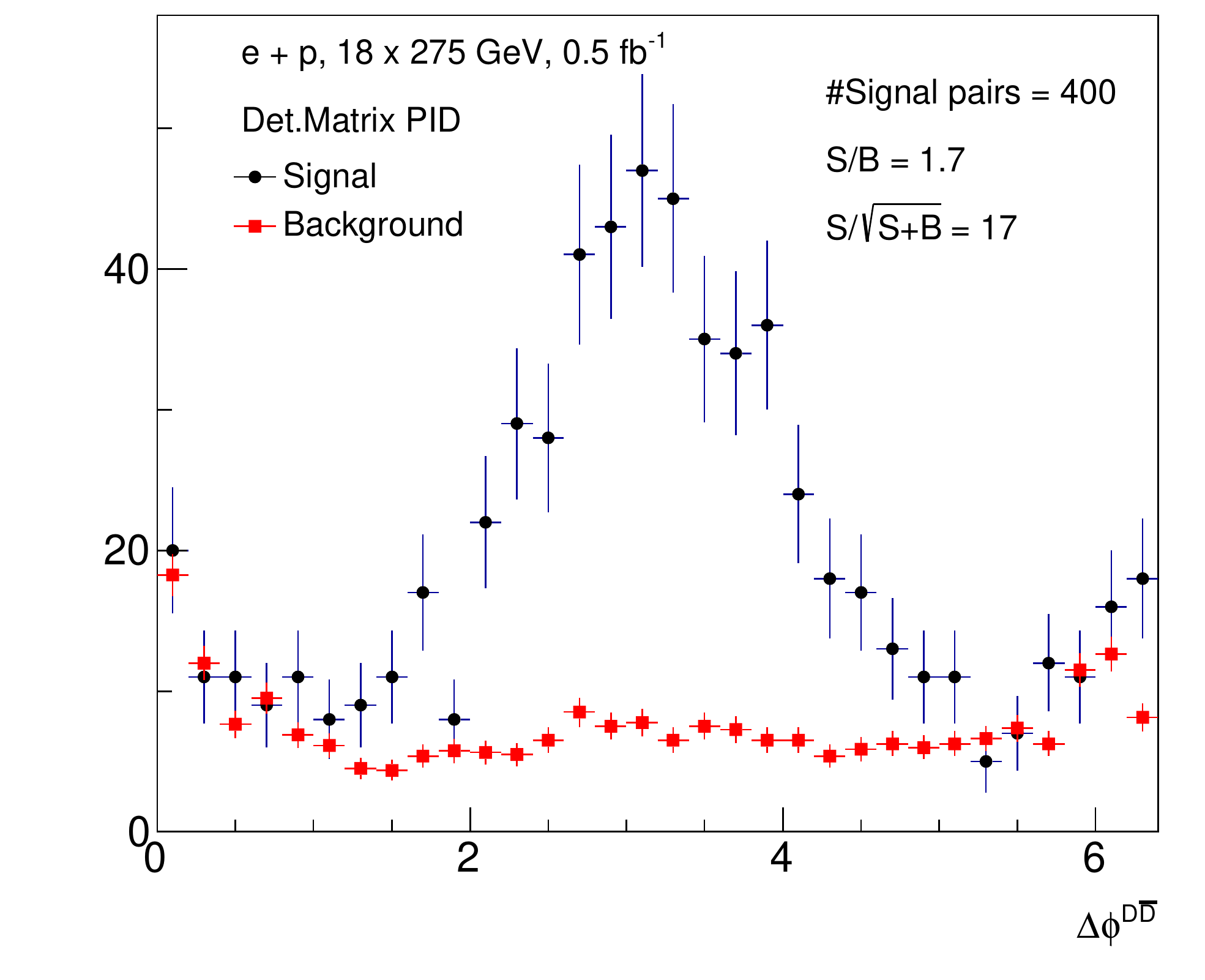}
}
   \caption{The azimuthal angle difference between $D^{0}$ and $\overline{D^{0}}$ candidate pairs in an event for candidates within a 3$\sigma$ mass window around the $D^{0}$ mass peak (Signal) and for candidates between 6$\sigma$ and 12$\sigma$ outside the mass peak on either side (Background) for e + p collisions at beam energies 18 x 275 GeV. The number of Signal $D^{0}-\overline{D^{0}}$ pairs, signal to background ratio and signal significance are also indicated. 
}
\label{fig:pairrecoexp}
\end{figure}

The explicit reconstruction of heavy flavor hadron pairs is studied by correlating the reconstructed $D^0$ and $\overline{D^0}$ candidates in the event. Figure~\ref{fig:pairrecoexp} shows the azimuthal angle difference $\Delta\phi$ between the reconstructed momenta of $D^0$ and $\overline{D^0}$ candidates in the same event. The signal distribution is from $D^0$ and $\overline{D^0}$ candidate pairs within 3$\sigma$ mass window of the nominal $D^0$ mass. The background is constructed using $D^0$ and $\overline{D^0}$ candidate pairs that are within 6$\sigma$ and 12$\sigma$ on either side of the mass window, normalized to the same mass window width and averaged between the different pair combinations for combining candidate pairs on the two sides of the mass window. The same topological variable selection cuts for $D^0$ and $\overline{D^0}$ candidate selection described in the previous section are applied in the reconstruction. The explicit reconstruction gives a clean signal with small background and allows an exact subtraction of the background. However, it relies on having good PID capabilities both in the forward and backward regions. The signal significance, $\sigma_{\mathrm{N}} = S/\sqrt{S+B}$, with $S$ and $B$ being the signal and background counts, corresponding to the generated luminosity is also indicated in the figure.  

The explicit reconstruction of heavy flavor hadrons suffer from small branching ratios to the hadronic channels, for e.g. the $D^0 \rightarrow K^- \pi^+$ branching ratio is only 3.89\%~\cite{Zyla:2020zbs}. A possible way to overcome this loss of statistics from the poor branching ratio is to tag the heavy flavor hadron decays utilizing the features of their decay topology. The tracker with excellent track pointing resolution allows tagging of their decay vertices. The downside is that tagging won't give pure signals like in the case of explicit reconstruction, and the selection cuts need to be optimized to attain the best purity and signal significance. 

\begin{figure}[htbp]
\center{
\includegraphics[width=1.0\columnwidth]{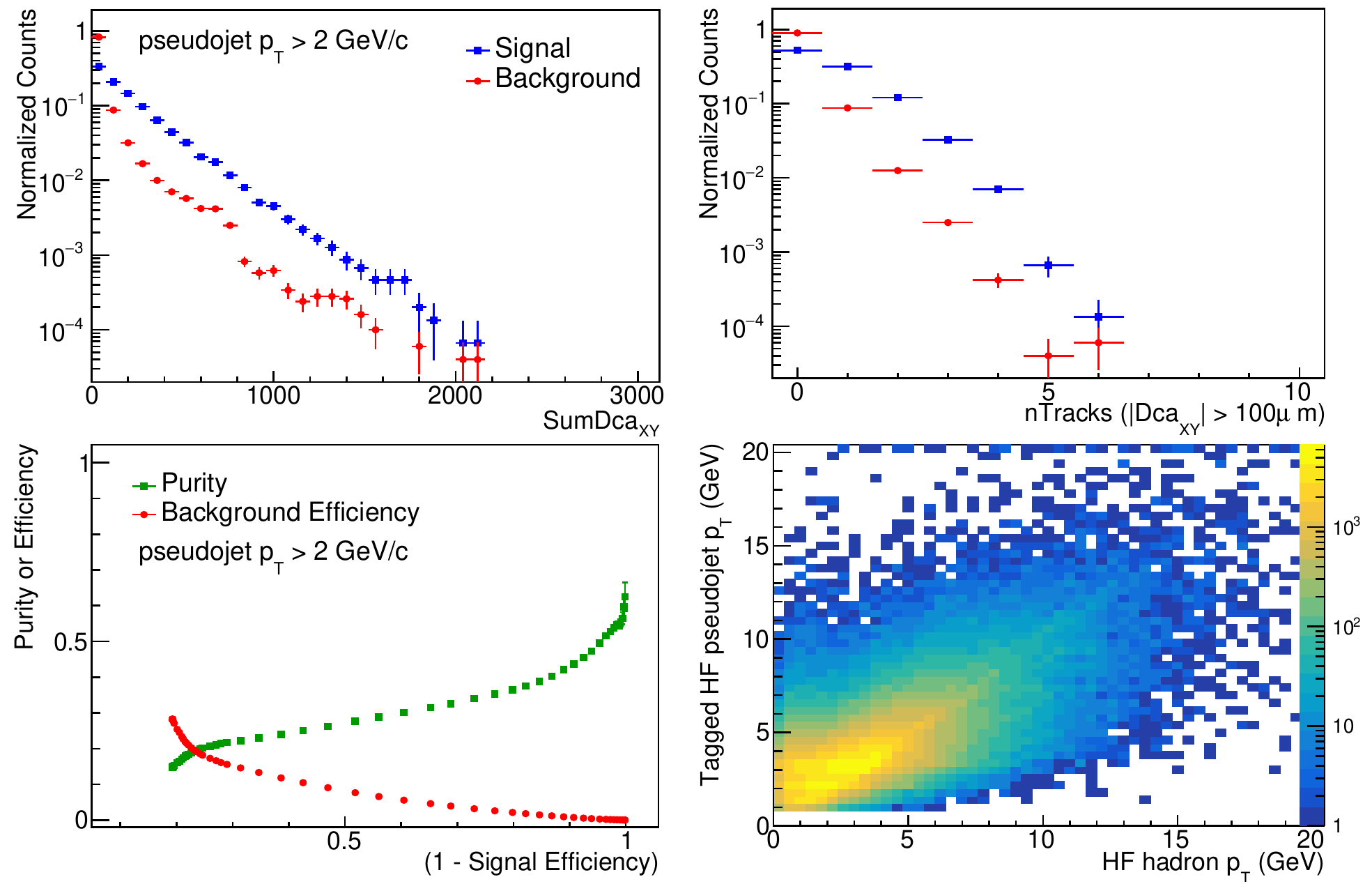}
}
   \caption{ The normalized distributions of summed $|DCA_{\mathrm{XY}}|$ of tracks within a tagged pseudojet (top left), number of tracks with a $|DCA_{\mathrm{XY}}|$ of at least 100 $\mu$m from the primary vertex within a tagged pseudojet (top right), the signal purity for and background (light flavor) efficiency for heavy flavor tagging for reconstructed pseudojets with $p_{\mathrm{T}} >$ 2 GeV/c, and (bottom right) correlation between $p_{\mathrm{T}}$ of parent heavy-flavor hadron and the reconstructed tagged pseudojet matched to the heavy-flavor hadron.
}
\label{fig:tagvars}
\end{figure}

In order to perform tagging, pseudojets are defined using the anti-$k_{\mathrm{T}}$ clustering algorithm~\cite{Cacciari:2008gp} provided by FASTJET package~\cite{Cacciari:2011ma}. The track candidates going into the clustering algorithm are required to have a minimum $p_{\mathrm{T}}$ of 0.2 GeV/c and to be within detector acceptance $|\eta| <$ 3.0. A pseudojet radius $\Delta R$ = 1.0 is chosen for clustering. All clusters in an event returned by the clustering algorithm are taken as pseudojets in the analysis, and the momentum of the cluster is taken as the pseudojet momentum. To better isolate heavy flavor pseudojets, we utilize a few variables characterizing their decay topology features. The sum of absolute values of DCA of all tracks associated with pseudojet (sumDCA), the number of tracks associated with the pseudojet with a minimum displacement of 100 $\mu$m (nTracks) from the PV and the minimum DCA between displaced (minimum DCA of 100 $\mu$m from the PV) track pairs within the pseudojet are found to have good differentiation between heavy flavor and light flavor pseudojets. The sumDCA is evaluated with a maximum DCA cut of $|DCA| <$ 750 $\mu$m for tracks to reduce contribution from strange hadron decays. The first two variables give the best separation and are shown in the top panels of Fig~\ref{fig:tagvars}. The signal are pseudojets matched to heavy flavor (includes both charm and bottom) hadrons and background are those matched to light flavor hadrons. The truth matching is done by associating a pseudojet to a heavy flavor hadron, if the parent hadron momentum vector falls within the pseudojet cone. If more than one heavy flavor hadrons do so, the one closest to the jet axis is taken. Pseudojets without heavy flavor hadrons are tagged light flavor pseudojets. The different variables are combined using a Boosted Decision Tree (BDT) algorithm provided by the TMVA package~\cite{Hocker:2007ht} to give a single response that can be utilized for discriminating heavy flavor and light flavor pseudojets. The BDT is trained using an independent signal sample of heavy flavor hadron pseudojets and a background sample of light flavor pseudojets produced using the same simulation setup as used for the analysis. Before the BDT selection and training a simple cut of sumDCA $>$ 50 $\mu$m is applied without much loss of signal efficiency. 

The heavy flavor tagging performance is shown in the lower left panel of Fig~\ref{fig:tagvars}. The purity and background selection efficiencies are shown as a function of the signal efficiency. The different points shown are with different selection cuts on the BDT response variable. The signal purity without any selection cuts is 2\% (not shown on the plot). With topological selection, a signal purity of about 50\% can be achieved with signal efficiency of $\sim$10\% and a purity of about 60\% with close to tightest cuts on signal efficiency. The bottom right panel shows the correlation between the $p_{\mathrm{T}}$ of the parent heavy flavor hadron and that of the matched pseudojet. Good correlation is seen between the two, with some smearing. This shows the kinematics of the parent hadron can still be accessed even without the full explicit reconstruction. 

\begin{figure}[htbp]
\center{
\includegraphics[width=1.0\columnwidth]{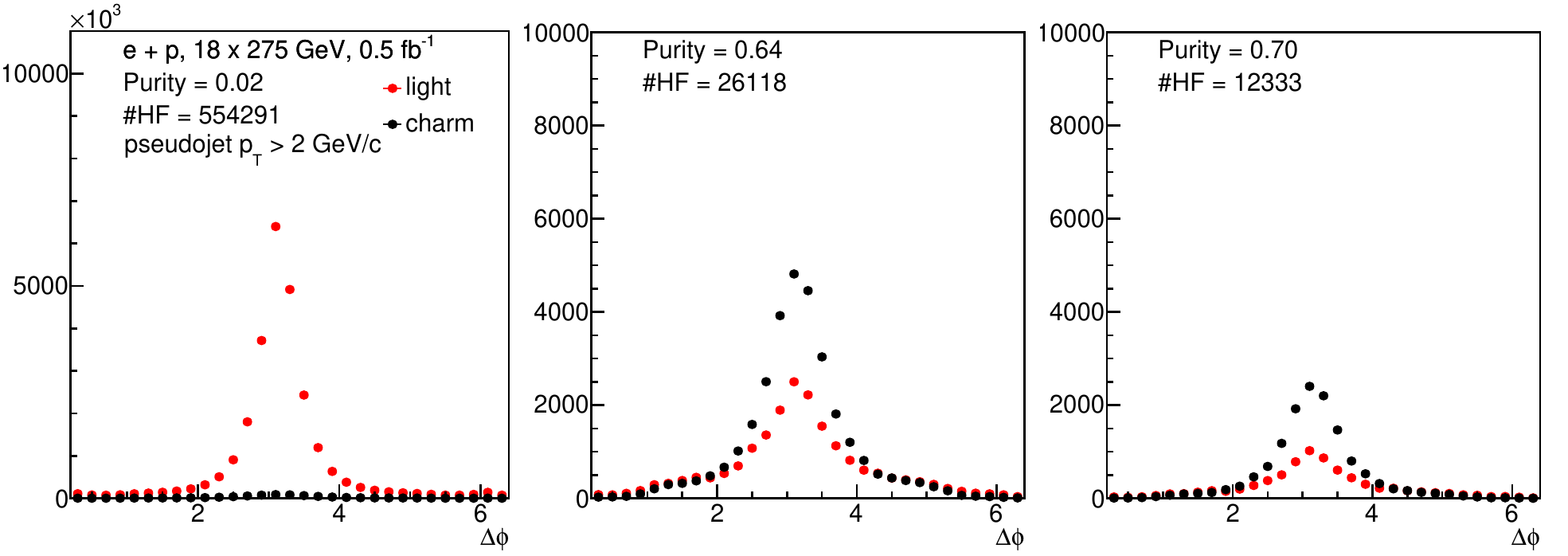}
}  
   \caption{ The azimuthal angle difference between tagged pseudojet pairs in an event for light flavor (solid circles) and heavy flavor (solid squares) pseudojets without the topological selection (left) and with the selection (middle, right) for e + p collisions at beam energies 18 x 275 GeV. Pseudojets with $p_{\mathrm{T}} >$ 2 GeV/c are used in constructing the pairs. The purity, evaluated in 2.0 $< |\Delta\phi| < $ 4.4,for heavy flavor selection indicated in the panels.
}
\label{fig:recopairtag}
\end{figure}

Tagged heavy flavor pair pairs are constructed using the correlation between psuedojets in the same event, similar to that for explicit reconstruction. Figure~\ref{fig:recopairtag} shows the $\Delta\phi$ distributions between the azimuthal angle of the momenta of the pseudojets in an event for the case without any selection on the decay topology and with two different selections. The number of signal pairs as well as the signal purity for the generated luminosity are also shown on the plots. The topological selection allows to improve the signal purity to $~$70\% (from 2\%) at a reasonable signal efficiency. 

\begin{figure}[htbp]
\center{
\includegraphics[width=0.8\columnwidth]{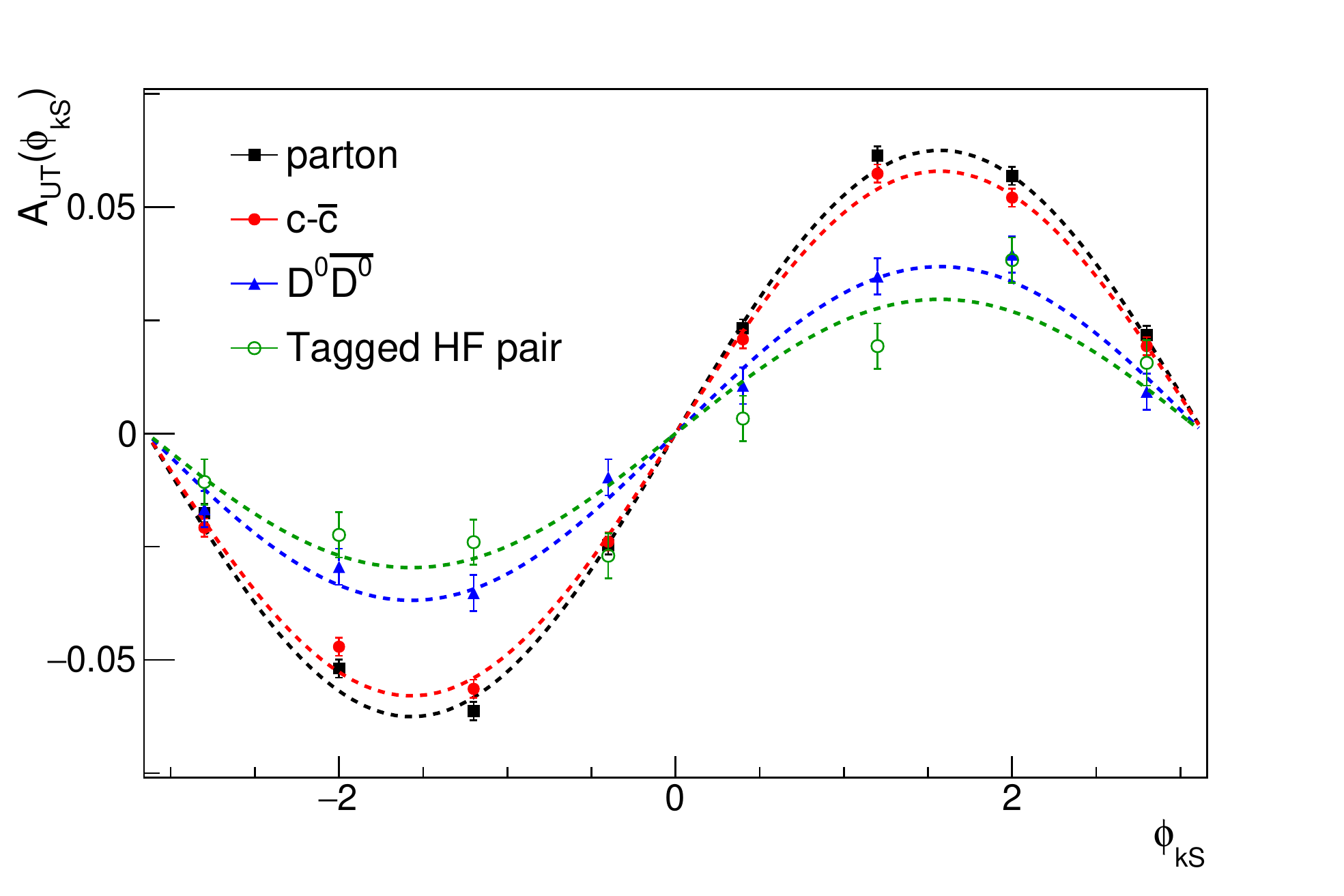}
}
   \caption{ The transverse asymmetry $A_{\mathrm{UT}}$ input at the parton level (solid squares), reconstructed from the $c-\bar{c}$ pair after the hard-scattering (solid circles), reconstructed from the $D^0-\overline{D^0}$ pair (solid triangles) and from the tagged heavy flavor pairs (open circles) evaluated using events generated with PYTHIA 6.4, shown as a function of the azimuthal angle corresponding to the gluon/heavy quark pair/heavy hadron pair momentum ($\phi_{\mathrm{kS}}$). The photon momentum is subtracted in evaluating the heavy quark/hadron pair momentum. The error bars are not scaled to luminosity.
}
\label{fig:decor}
\end{figure}

We have also evaluated the correlations between the asymmetry at the gluon level and the corresponding asymmetries at the reconstructed levels, for both the explicit reconstruction of $D^0\overline{D^0}$ pair and also tagged heavy flavor hadron pairs. The decorrelation of the signal at the gluon level and different stages of the scattering, hadronization and reconstruction are shown in Figure~\ref{fig:decor}. The evaluation is done using events generated with PYTHIA 6.4. For this evaluation, a constant input asymmetry at gluon level (6\%) is input by hand, irrespective of the gluon kinematics, by modulating the azimuthal angle corresponding to the gluon transverse momentum. The azimuthal angle distributions of the transverse momenta of the heavy quark pairs after hard scattering and the reconstructed (and tagged) heavy flavor hadrons after hadronization are then calculated to evaluate the loss of input signal at each stage. The signal reduces by about 30\% at hadronization, the tagging reduces the signal bit more, reducing to about 50\% of the parton level signal. Much of the signal and correlation is still retained in either case and can serve as probes to study the gluon TMD.

\section{Projections for asymmetry measurements}
\label{sec:proj}

Statistical uncertainty projection for a measurement of the transverse asymmetry $A_{\mathrm{UT}}$ can be made in a straight forward way. For reconstructed $D^0 - \overline{D^0}$ pairs, from eq.~\ref{eq1}, the uncertainty is evaluated as $\delta A_{\mathrm{UT}} = \sqrt(\frac{1}{P^2N} - A_{\mathrm{UT}}^2/N)$, where $A_\mathrm{UT}$ is the magnitude of the signal, $N$ is the number of $D^0 - \overline{D^0}$ pairs in the sample which is evaluated from the signal significance $\sigma_{\mathrm{N}}$ as $N = \sigma_{\mathrm{N}}^2$ and $P$ is proton beam polarization. The second term is much smaller compared to the first, even if a relatively large value for $A_{\mathrm{UT}}$ is assumed. A similar calculation holds for the case of the $\langle \cos{(2\phi_{\mathrm{T}})} \rangle$ observable corresponding to the linearly polarized gluon TMD, except that the reduction in precision from beam polarization would not be there, as these can be measured in unpolarized $ep$ collisions.  

The $A_{\mathrm{UT}}$ signal from tagged pseudojet pairs is a weighted sum of both heavy flavor and light flavor $A_{\mathrm{UT}}$ values, $A_{\mathrm{UT}}^{meas} = p_{\mathrm{h}}A_{\mathrm{UT}}^{\mathrm{h}} + p_{\mathrm{l}}A_{\mathrm{UT}}^{\mathrm{l}}$. Here $p_{\mathrm{h}}$ ($p_{\mathrm{l}}$) denote the purity, defined as the fraction of heavy (light) flavor pseudojets in the sample. With and without the topological selection (tagging), the purities can be altered significantly. And without going into the exact values of the purities, just from their general magnitudes as shown in Fig~\ref{fig:recopairtag}, it can be shown that $A_{\mathrm{UT}}^{h} = (A_{\mathrm{UT}}^{meas,s} - p_{\mathrm{l}}^{\mathrm{s}}A_{\mathrm{UT}}^{\mathrm{meas,0}})/p_{\mathrm{h}}^{\mathrm{s}}$. The superscripts $s$ and $0$ denote the samples with and without the topological selections. Since $A_{\mathrm{UT}}^{meas,0}$ has far higher precision, and taking into account the finite beam polarization $P$, we have for the statistical uncertainty corresponding to the Sivers TMD $\delta A_{\mathrm{UT}}^{h} =  \delta A_{\mathrm{UT}}^{\mathrm{meas,s}}/p_{\mathrm{h}}^{\mathrm{s}}/P$, with the quantity $\delta A_{\mathrm{UT}}^{\mathrm{meas,s}}$ denoting the uncertainty on the measured $A_{\mathrm{UT}}^{meas,s}$ with tagged pseudojets. Again, a similar expression without the dilution from beam polarization, would give the uncertainty projection for the $\langle \cos{(2\phi_{\mathrm{T}})} \rangle$ observable. 

\begin{figure*}[htbp]
\center{
\includegraphics[width=1.9\columnwidth]{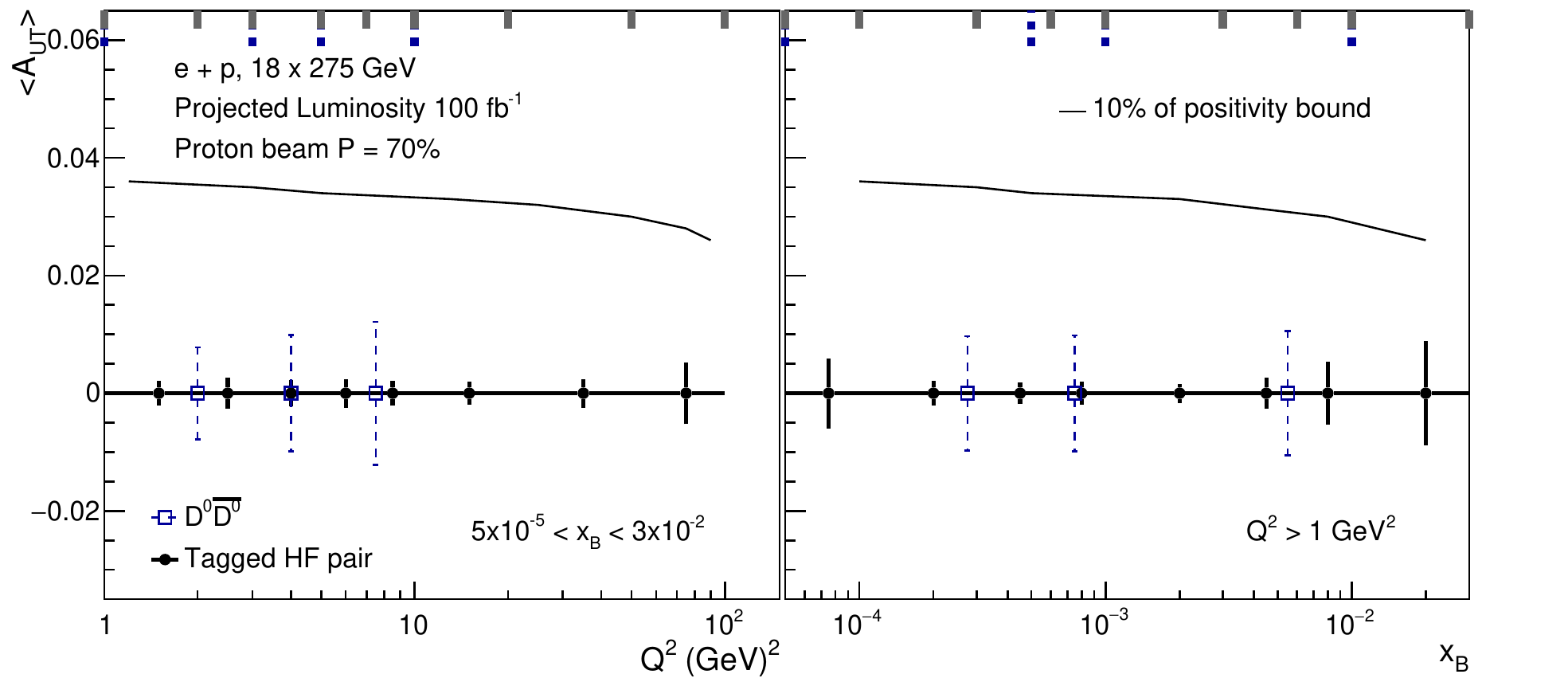}
}  
   \caption{ Projected statistical uncertainties for $A_{\mathrm{UT}}^{\mathrm{h}}$ of heavy flavor hadron pairs, shown as error bars around zero, in different $Q^2$ (left) and $x_{\mathrm{B}}$ (right) bins using exclusively reconstructed $D^0-\overline{D^0}$ pairs (dashed lines) and tagged heavy flavor pairs (solid lines). The projections are shown for a luminosity of 100 fb$^{\mathrm{-1}}$ e + p collisions at beam energies 18 $\times$ 275 GeV. The bin boundaries corresponding to the projections for the exclusive and tagged cases are indicated by dashed and solid lines on the top axis, respectively. The solid curve shows the $A_{\mathrm{UT}}$ values for $D^0 - \overline{D^0}$ pairs evaluated through PYTHIA 6.4 with an input gluon Sivers effect equal to 10\% of the positivity bound.
}
\label{fig:proj}
\end{figure*}

Figure~\ref{fig:proj} shows the statistical uncertainty projections in different $Q^2$ and $x_{\mathrm{B}}$ bins from explicit reconstruction of $D^0 - \overline{D^0}$ pairs and tagged heavy flavor pairs for the transverse asymmetry $A_{\mathrm{UT}}^{\mathrm{h}}$. With a proton beam polarization of 70$\%$, and a projected integrated luminosity for polarized p + e collisions at the EIC of 100 fb$^{\mathrm{-1}}$, the absolute statistical uncertainty on $A_{\mathrm{UT}}$ is 0.58\% for $Q^2 >$ 1 GeV$^{\mathrm{2}}$ and $\mathrm{5\times10}^{-\mathrm{5}} < x_{\mathrm{B}} < \mathrm{10}^{-\mathrm{2}}$, with reconstructed $D^0 - \overline{D^0}$ pairs. The tagging improves the uncertainties on $A_{\mathrm{UT}}^{\mathrm{h}}$ measurements significantly. The uncertainty for $Q^{2} >$ 1 GeV$^2$ and $\mathrm{5\times10}^{-\mathrm{4}} < x_{\mathrm{B}} < \mathrm{10}^{-\mathrm{2}}$ reduces to 0.08\%, by about a factor of 7 with tagged heavy flavor pairs. The measurements would also offer extended kinematic reach in both $Q^2$ and $x_{\mathrm{B}}$ for the gluon TMD measurements with good precision. The tagging also offers the advantage that changes in momentum resolution (from the choice of the magnetic field for the experiment) does not directly reduce the performance, unlike in the case of explicit reconstruction where the mass peak broadens. The tagging algorithm studied here is solely utilizing the track pointing resolution capabilities of the detector (in the transverse direction) and does not depend on PID requirements in the forward or backward directions. However, incorporating PID into tagging, could help improve the performance. So too would utilizing the track pointing capabilities in the $z$ direction. These are left for future studies that can be explored with the detector designs and specifications at a more mature stage.

Compared to the predicted magnitudes of $\langle \cos{(2\phi_{\mathrm{T}})} \rangle$, the projected uncertainties are much smaller, allowing for good precision measurements. For the gluon Sivers TMD also these measurements will offer constraints to lower $x$ and higher $Q^2$. The $A_{\mathrm{UT}}$ values for $D^0 - \overline{D^0}$ pairs evaluated through PYTHIA 6.4 with an input gluon Sivers TMD equal to 10\% of the positivity bound and quark Sivers TMD from the SIDIS fits~\cite{Anselmino:2016uie}, is also shown in Fig~\ref{fig:proj}. The partons initiating the processes were given transverse asymmetries corresponding to the respective Sivers TMD at the $x$ and $Q^2$ values of the process, and the resulting $A_{\mathrm{UT}}$ values for $D^0 - \overline{D^0}$ pairs evaluated, following the procedure in ~\cite{Zheng:2018ssm}. Although there aren't strong constraints on the gluon Sivers TMD, some of the recent estimates put it at a few percent of the positivity bound~\cite{DAlesio:2015fwo,DAlesio:2018rnv}. Another attractive channel to study the gluon transverse asymmetries is the di-jet $A_{\mathrm{UT}}$ measurements. The projections for di-jet $A_{\mathrm{UT}}$ uncertainties are a few times (for eg., about 4 times in the bin $\mathrm{10}^{-\mathrm{3}} < x_{\mathrm{B}} < \mathrm{3\times10}^{-\mathrm{3}}$, $Q^2 >$ 1 GeV$^{\mathrm{2}}$) better than for the tagged heavy flavor hadron pair $A_{\mathrm{UT}}$. However, the heavy flavor hadron pair measurements offers a more sensitive and direct probe of the gluon asymmetries, as the di-jet measurements get sizeable contributions from the quark asymmetries as well. The tagged heavy flavor hadron pair channel therefore offers a complementary and attractive channel to measure the gluon TMD. 

\begin{figure}[htbp]
\includegraphics[width=0.95\columnwidth]{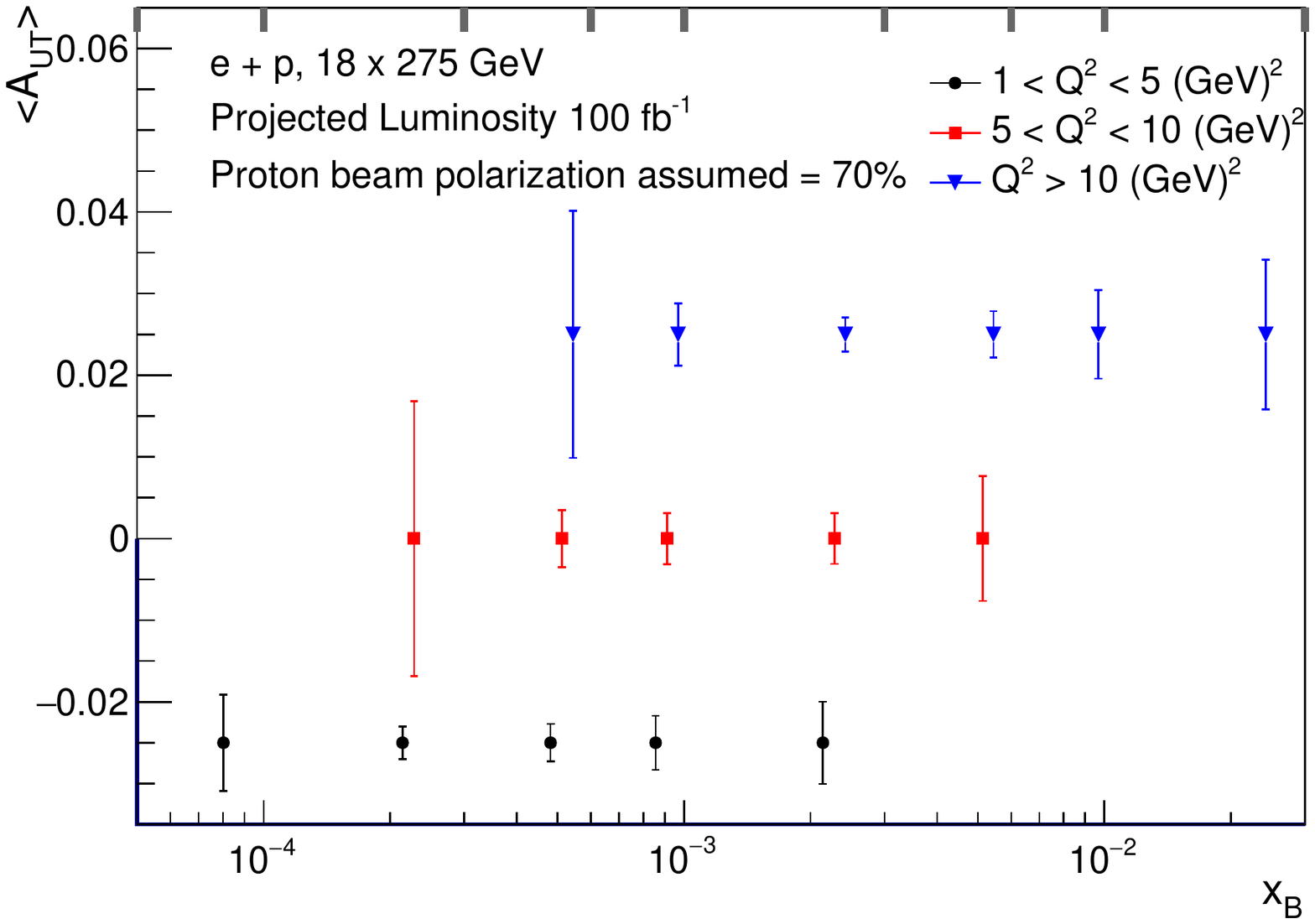}
   \caption{ Projected statistical uncertainties for $A_{\mathrm{UT}}^{\mathrm{h}}$, shown as error bars, in different $x_{\mathrm{B}}$ bins for different $Q^2$ selections using tagged heavy flavor pairs. The $A_{\mathrm{UT}}^{\mathrm{h}}$ values around which the error bars are shown are arbitrary and chosen for plotting convenience. The projections are shown for a luminosity of 100 fb$^{\mathrm{-1}}$ e + p collisions at beam energies 18 $\times$ 275 GeV. The bin boundaries in $x_{\mathrm{B}}$ corresponding to the projections are indicated by solid lines on the top axis. 
}
\label{fig:proj2}
\end{figure}

The statistical uncertainty projections for $A_{\mathrm{UT}}^{\mathrm{h}}$ using tagged heavy flavor pairs in more differential bins with simultaneous binning in both $Q^2$ and $x_{\mathrm{B}}$ are shown in Fig~\ref{fig:proj2}. The values of $A_{\mathrm{UT}}^{\mathrm{h}}$ shown are arbitrary, chosen for visibility, while the error bars give the projected statistical uncertainties. The uncertainties are at sub-percent level for most of the $Q^2$ and $x_{\mathrm{B}}$ bins, with higher $Q^2$ values giving access to higher $x_{\mathrm{B}}$ bins and vice-versa, reflecting the $x_{\mathrm{B}} - Q^2$ dependence of heavy flavor production cross-section in DIS events~\cite{Kelsey:2021gpk}.  We have also explored other beam energy configurations for $e$ and $p$ beams, particularly the lower energy configuration with 5 GeV electron and 100 GeV proton beams. The lower energy configuration was found to give larger projected uncertainties at all $x_{\mathrm{B}}$ values as the total charm production cross-section decreases.   

\section{Summary}
We have presented a study of heavy flavor hadron pair reconstruction performance at an EIC detector with MAPS based silicon tracker and vertexing subsystems, using events generated using  PYTHIA 6.4 simulations. Exclusive reconstruction of the heavy flavor hadron pairs via their hadronic decays provide an experimentally clean measurement, however suffers from loss of statistics due to small hadronic branching ratios of heavy flavor hadrons. A heavy flavor tagging algorithm is developed utilizing the decay topology of heavy flavor hadrons and the track pointing capabilities of the detector. The tagging is found to improve the statistical precision of heavy flavor measurements significantly with good purity for heavy flavor selection. The topological tagging enhances the purity for heavy flavor selection by a factor of 35, compared to without the topological selection. The initial azimuthal asymmetry in gluon distributions were found to be retained, with some dilution, by the final state heavy flavor hadron pair, for both exclusive reconstruction and tagged heavy flavor hadrons. Statistical uncertainty projections for the heavy flavor hadron pair transverse asymmetries corresponding to the gluon Sivers TMD (and TMD of linearly polarized gluons) are evaluated. The heavy flavor hadron tagging is found to give about an order of magnitude improved uncertainty projections compared to using exclusive reconstruction of $D^0 - \overline{D^0}$ pairs. The tagged heavy flavor pair measurements can provide a gluon rich measurement compared to inclusive di-hadrons and di-jets measurements. This study opens up heavy flavor hadron pair measurements as an attractive, complementary and independent channel to access gluon TMD at the EIC. 

\bibliography{refs}

\begin{thebibliography}{66}%
\makeatletter
\providecommand \@ifxundefined [1]{%
 \@ifx{#1\undefined}
}%
\providecommand \@ifnum [1]{%
 \ifnum #1\expandafter \@firstoftwo
 \else \expandafter \@secondoftwo
 \fi
}%
\providecommand \@ifx [1]{%
 \ifx #1\expandafter \@firstoftwo
 \else \expandafter \@secondoftwo
 \fi
}%
\providecommand \natexlab [1]{#1}%
\providecommand \enquote  [1]{``#1''}%
\providecommand \bibnamefont  [1]{#1}%
\providecommand \bibfnamefont [1]{#1}%
\providecommand \citenamefont [1]{#1}%
\providecommand \href@noop [0]{\@secondoftwo}%
\providecommand \href [0]{\begingroup \@sanitize@url \@href}%
\providecommand \@href[1]{\@@startlink{#1}\@@href}%
\providecommand \@@href[1]{\endgroup#1\@@endlink}%
\providecommand \@sanitize@url [0]{\catcode `\\12\catcode `\$12\catcode
  `\&12\catcode `\#12\catcode `\^12\catcode `\_12\catcode `\%12\relax}%
\providecommand \@@startlink[1]{}%
\providecommand \@@endlink[0]{}%
\providecommand \url  [0]{\begingroup\@sanitize@url \@url }%
\providecommand \@url [1]{\endgroup\@href {#1}{\urlprefix }}%
\providecommand \urlprefix  [0]{URL }%
\providecommand \Eprint [0]{\href }%
\providecommand \doibase [0]{http://dx.doi.org/}%
\providecommand \selectlanguage [0]{\@gobble}%
\providecommand \bibinfo  [0]{\@secondoftwo}%
\providecommand \bibfield  [0]{\@secondoftwo}%
\providecommand \translation [1]{[#1]}%
\providecommand \BibitemOpen [0]{}%
\providecommand \bibitemStop [0]{}%
\providecommand \bibitemNoStop [0]{.\EOS\space}%
\providecommand \EOS [0]{\spacefactor3000\relax}%
\providecommand \BibitemShut  [1]{\csname bibitem#1\endcsname}%
\let\auto@bib@innerbib\@empty
\bibitem [{\citenamefont {Grosse~Perdekamp}\ and\ \citenamefont
  {Yuan}(2015)}]{GrossePerdekamp:2015xdx}%
  \BibitemOpen
  \bibfield  {author} {\bibinfo {author} {\bibfnamefont {M.}~\bibnamefont
  {Grosse~Perdekamp}}\ and\ \bibinfo {author} {\bibfnamefont {F.}~\bibnamefont
  {Yuan}},\ }\href {\doibase 10.1146/annurev-nucl-102014-021948} {\bibfield
  {journal} {\bibinfo  {journal} {Ann. Rev. Nucl. Part. Sci.}\ }\textbf
  {\bibinfo {volume} {65}},\ \bibinfo {pages} {429} (\bibinfo {year} {2015})},\
  \Eprint {http://arxiv.org/abs/1510.06783} {arXiv:1510.06783 [hep-ph]}
  \BibitemShut {NoStop}%
\bibitem [{\citenamefont {Sivers}(1990)}]{Sivers:1989cc}%
  \BibitemOpen
  \bibfield  {author} {\bibinfo {author} {\bibfnamefont {D.~W.}\ \bibnamefont
  {Sivers}},\ }\href {\doibase 10.1103/PhysRevD.41.83} {\bibfield  {journal}
  {\bibinfo  {journal} {Phys. Rev. D}\ }\textbf {\bibinfo {volume} {41}},\
  \bibinfo {pages} {83} (\bibinfo {year} {1990})}\BibitemShut {NoStop}%
\bibitem [{\citenamefont {Boer}\ \emph {et~al.}(2011)\citenamefont {Boer},
  \citenamefont {Brodsky}, \citenamefont {Mulders},\ and\ \citenamefont
  {Pisano}}]{Boer:2010zf}%
  \BibitemOpen
  \bibfield  {author} {\bibinfo {author} {\bibfnamefont {D.}~\bibnamefont
  {Boer}}, \bibinfo {author} {\bibfnamefont {S.~J.}\ \bibnamefont {Brodsky}},
  \bibinfo {author} {\bibfnamefont {P.~J.}\ \bibnamefont {Mulders}}, \ and\
  \bibinfo {author} {\bibfnamefont {C.}~\bibnamefont {Pisano}},\ }\href
  {\doibase 10.1103/PhysRevLett.106.132001} {\bibfield  {journal} {\bibinfo
  {journal} {Phys. Rev. Lett.}\ }\textbf {\bibinfo {volume} {106}},\ \bibinfo
  {pages} {132001} (\bibinfo {year} {2011})},\ \Eprint
  {http://arxiv.org/abs/1011.4225} {arXiv:1011.4225 [hep-ph]} \BibitemShut
  {NoStop}%
\bibitem [{\citenamefont {Collins}(2013)}]{Collins:2011zzd}%
  \BibitemOpen
  \bibfield  {author} {\bibinfo {author} {\bibfnamefont {J.}~\bibnamefont
  {Collins}},\ }\href@noop {} {\emph {\bibinfo {title} {{Foundations of
  perturbative QCD}}}},\ Vol.~\bibinfo {volume} {32}\ (\bibinfo  {publisher}
  {Cambridge University Press},\ \bibinfo {year} {2013})\BibitemShut {NoStop}%
\bibitem [{\citenamefont {Adams}\ \emph {et~al.}(1991)\citenamefont {Adams}
  \emph {et~al.}}]{E581:1991eys}%
  \BibitemOpen
  \bibfield  {author} {\bibinfo {author} {\bibfnamefont {D.~L.}\ \bibnamefont
  {Adams}} \emph {et~al.} (\bibinfo {collaboration} {E581, E704}),\ }\href
  {\doibase 10.1016/0370-2693(91)91351-U} {\bibfield  {journal} {\bibinfo
  {journal} {Phys. Lett. B}\ }\textbf {\bibinfo {volume} {261}},\ \bibinfo
  {pages} {201} (\bibinfo {year} {1991})}\BibitemShut {NoStop}%
\bibitem [{\citenamefont {Krueger}\ \emph {et~al.}(1999)\citenamefont {Krueger}
  \emph {et~al.}}]{Krueger:1998hz}%
  \BibitemOpen
  \bibfield  {author} {\bibinfo {author} {\bibfnamefont {K.}~\bibnamefont
  {Krueger}} \emph {et~al.},\ }\href {\doibase 10.1016/S0370-2693(99)00677-2}
  {\bibfield  {journal} {\bibinfo  {journal} {Phys. Lett. B}\ }\textbf
  {\bibinfo {volume} {459}},\ \bibinfo {pages} {412} (\bibinfo {year}
  {1999})}\BibitemShut {NoStop}%
\bibitem [{\citenamefont {Allgower}\ \emph {et~al.}(2002)\citenamefont
  {Allgower} \emph {et~al.}}]{Allgower:2002qi}%
  \BibitemOpen
  \bibfield  {author} {\bibinfo {author} {\bibfnamefont {C.~E.}\ \bibnamefont
  {Allgower}} \emph {et~al.},\ }\href {\doibase 10.1103/PhysRevD.65.092008}
  {\bibfield  {journal} {\bibinfo  {journal} {Phys. Rev. D}\ }\textbf {\bibinfo
  {volume} {65}},\ \bibinfo {pages} {092008} (\bibinfo {year}
  {2002})}\BibitemShut {NoStop}%
\bibitem [{\citenamefont {Adams}\ \emph {et~al.}(2004)\citenamefont {Adams}
  \emph {et~al.}}]{STAR:2003lxu}%
  \BibitemOpen
  \bibfield  {author} {\bibinfo {author} {\bibfnamefont {J.}~\bibnamefont
  {Adams}} \emph {et~al.} (\bibinfo {collaboration} {STAR}),\ }\href {\doibase
  10.1103/PhysRevLett.92.171801} {\bibfield  {journal} {\bibinfo  {journal}
  {Phys. Rev. Lett.}\ }\textbf {\bibinfo {volume} {92}},\ \bibinfo {pages}
  {171801} (\bibinfo {year} {2004})},\ \Eprint
  {http://arxiv.org/abs/hep-ex/0310058} {arXiv:hep-ex/0310058} \BibitemShut
  {NoStop}%
\bibitem [{\citenamefont {Adler}\ \emph {et~al.}(2005)\citenamefont {Adler}
  \emph {et~al.}}]{PHENIX:2005jxc}%
  \BibitemOpen
  \bibfield  {author} {\bibinfo {author} {\bibfnamefont {S.~S.}\ \bibnamefont
  {Adler}} \emph {et~al.} (\bibinfo {collaboration} {PHENIX}),\ }\href
  {\doibase 10.1103/PhysRevLett.95.202001} {\bibfield  {journal} {\bibinfo
  {journal} {Phys. Rev. Lett.}\ }\textbf {\bibinfo {volume} {95}},\ \bibinfo
  {pages} {202001} (\bibinfo {year} {2005})},\ \Eprint
  {http://arxiv.org/abs/hep-ex/0507073} {arXiv:hep-ex/0507073} \BibitemShut
  {NoStop}%
\bibitem [{\citenamefont {Abelev}\ \emph {et~al.}(2008)\citenamefont {Abelev}
  \emph {et~al.}}]{STAR:2008ixi}%
  \BibitemOpen
  \bibfield  {author} {\bibinfo {author} {\bibfnamefont {B.~I.}\ \bibnamefont
  {Abelev}} \emph {et~al.} (\bibinfo {collaboration} {STAR}),\ }\href {\doibase
  10.1103/PhysRevLett.101.222001} {\bibfield  {journal} {\bibinfo  {journal}
  {Phys. Rev. Lett.}\ }\textbf {\bibinfo {volume} {101}},\ \bibinfo {pages}
  {222001} (\bibinfo {year} {2008})},\ \Eprint {http://arxiv.org/abs/0801.2990}
  {arXiv:0801.2990 [hep-ex]} \BibitemShut {NoStop}%
\bibitem [{\citenamefont {Arsene}\ \emph {et~al.}(2008)\citenamefont {Arsene}
  \emph {et~al.}}]{BRAHMS:2008doi}%
  \BibitemOpen
  \bibfield  {author} {\bibinfo {author} {\bibfnamefont {I.}~\bibnamefont
  {Arsene}} \emph {et~al.} (\bibinfo {collaboration} {BRAHMS}),\ }\href
  {\doibase 10.1103/PhysRevLett.101.042001} {\bibfield  {journal} {\bibinfo
  {journal} {Phys. Rev. Lett.}\ }\textbf {\bibinfo {volume} {101}},\ \bibinfo
  {pages} {042001} (\bibinfo {year} {2008})},\ \Eprint
  {http://arxiv.org/abs/0801.1078} {arXiv:0801.1078 [nucl-ex]} \BibitemShut
  {NoStop}%
\bibitem [{\citenamefont {Adamczyk}\ \emph
  {et~al.}(2012{\natexlab{a}})\citenamefont {Adamczyk} \emph
  {et~al.}}]{STAR:2012hth}%
  \BibitemOpen
  \bibfield  {author} {\bibinfo {author} {\bibfnamefont {L.}~\bibnamefont
  {Adamczyk}} \emph {et~al.} (\bibinfo {collaboration} {STAR}),\ }\href
  {\doibase 10.1103/PhysRevD.86.032006} {\bibfield  {journal} {\bibinfo
  {journal} {Phys. Rev. D}\ }\textbf {\bibinfo {volume} {86}},\ \bibinfo
  {pages} {032006} (\bibinfo {year} {2012}{\natexlab{a}})},\ \Eprint
  {http://arxiv.org/abs/1205.2735} {arXiv:1205.2735 [nucl-ex]} \BibitemShut
  {NoStop}%
\bibitem [{\citenamefont {Adamczyk}\ \emph
  {et~al.}(2012{\natexlab{b}})\citenamefont {Adamczyk} \emph
  {et~al.}}]{STAR:2012ljf}%
  \BibitemOpen
  \bibfield  {author} {\bibinfo {author} {\bibfnamefont {L.}~\bibnamefont
  {Adamczyk}} \emph {et~al.} (\bibinfo {collaboration} {STAR}),\ }\href
  {\doibase 10.1103/PhysRevD.86.051101} {\bibfield  {journal} {\bibinfo
  {journal} {Phys. Rev. D}\ }\textbf {\bibinfo {volume} {86}},\ \bibinfo
  {pages} {051101} (\bibinfo {year} {2012}{\natexlab{b}})},\ \Eprint
  {http://arxiv.org/abs/1205.6826} {arXiv:1205.6826 [nucl-ex]} \BibitemShut
  {NoStop}%
\bibitem [{\citenamefont {Bland}\ \emph {et~al.}(2015)\citenamefont {Bland}
  \emph {et~al.}}]{AnDY:2013att}%
  \BibitemOpen
  \bibfield  {author} {\bibinfo {author} {\bibfnamefont {L.~C.}\ \bibnamefont
  {Bland}} \emph {et~al.} (\bibinfo {collaboration} {AnDY}),\ }\href {\doibase
  10.1016/j.physletb.2015.10.001} {\bibfield  {journal} {\bibinfo  {journal}
  {Phys. Lett. B}\ }\textbf {\bibinfo {volume} {750}},\ \bibinfo {pages} {660}
  (\bibinfo {year} {2015})},\ \Eprint {http://arxiv.org/abs/1304.1454}
  {arXiv:1304.1454 [hep-ex]} \BibitemShut {NoStop}%
\bibitem [{\citenamefont {Adare}\ \emph
  {et~al.}(2014{\natexlab{a}})\citenamefont {Adare} \emph
  {et~al.}}]{PHENIX:2013wle}%
  \BibitemOpen
  \bibfield  {author} {\bibinfo {author} {\bibfnamefont {A.}~\bibnamefont
  {Adare}} \emph {et~al.} (\bibinfo {collaboration} {PHENIX}),\ }\href
  {\doibase 10.1103/PhysRevD.90.012006} {\bibfield  {journal} {\bibinfo
  {journal} {Phys. Rev. D}\ }\textbf {\bibinfo {volume} {90}},\ \bibinfo
  {pages} {012006} (\bibinfo {year} {2014}{\natexlab{a}})},\ \Eprint
  {http://arxiv.org/abs/1312.1995} {arXiv:1312.1995 [hep-ex]} \BibitemShut
  {NoStop}%
\bibitem [{\citenamefont {Adare}\ \emph
  {et~al.}(2014{\natexlab{b}})\citenamefont {Adare} \emph
  {et~al.}}]{PHENIX:2014qwb}%
  \BibitemOpen
  \bibfield  {author} {\bibinfo {author} {\bibfnamefont {A.}~\bibnamefont
  {Adare}} \emph {et~al.} (\bibinfo {collaboration} {PHENIX}),\ }\href
  {\doibase 10.1103/PhysRevD.90.072008} {\bibfield  {journal} {\bibinfo
  {journal} {Phys. Rev. D}\ }\textbf {\bibinfo {volume} {90}},\ \bibinfo
  {pages} {072008} (\bibinfo {year} {2014}{\natexlab{b}})},\ \Eprint
  {http://arxiv.org/abs/1406.3541} {arXiv:1406.3541 [hep-ex]} \BibitemShut
  {NoStop}%
\bibitem [{\citenamefont {Airapetian}\ \emph {et~al.}(2014)\citenamefont
  {Airapetian} \emph {et~al.}}]{HERMES:2013quo}%
  \BibitemOpen
  \bibfield  {author} {\bibinfo {author} {\bibfnamefont {A.}~\bibnamefont
  {Airapetian}} \emph {et~al.} (\bibinfo {collaboration} {HERMES}),\ }\href
  {\doibase 10.1016/j.physletb.2013.11.021} {\bibfield  {journal} {\bibinfo
  {journal} {Phys. Lett. B}\ }\textbf {\bibinfo {volume} {728}},\ \bibinfo
  {pages} {183} (\bibinfo {year} {2014})},\ \Eprint
  {http://arxiv.org/abs/1310.5070} {arXiv:1310.5070 [hep-ex]} \BibitemShut
  {NoStop}%
\bibitem [{\citenamefont {Airapetian}\ \emph {et~al.}(2005)\citenamefont
  {Airapetian} \emph {et~al.}}]{HERMES:2004mhh}%
  \BibitemOpen
  \bibfield  {author} {\bibinfo {author} {\bibfnamefont {A.}~\bibnamefont
  {Airapetian}} \emph {et~al.} (\bibinfo {collaboration} {HERMES}),\ }\href
  {\doibase 10.1103/PhysRevLett.94.012002} {\bibfield  {journal} {\bibinfo
  {journal} {Phys. Rev. Lett.}\ }\textbf {\bibinfo {volume} {94}},\ \bibinfo
  {pages} {012002} (\bibinfo {year} {2005})},\ \Eprint
  {http://arxiv.org/abs/hep-ex/0408013} {arXiv:hep-ex/0408013} \BibitemShut
  {NoStop}%
\bibitem [{\citenamefont {Qian}\ \emph {et~al.}(2011)\citenamefont {Qian} \emph
  {et~al.}}]{JeffersonLabHallA:2011ayy}%
  \BibitemOpen
  \bibfield  {author} {\bibinfo {author} {\bibfnamefont {X.}~\bibnamefont
  {Qian}} \emph {et~al.} (\bibinfo {collaboration} {Jefferson Lab Hall A}),\
  }\href {\doibase 10.1103/PhysRevLett.107.072003} {\bibfield  {journal}
  {\bibinfo  {journal} {Phys. Rev. Lett.}\ }\textbf {\bibinfo {volume} {107}},\
  \bibinfo {pages} {072003} (\bibinfo {year} {2011})},\ \Eprint
  {http://arxiv.org/abs/1106.0363} {arXiv:1106.0363 [nucl-ex]} \BibitemShut
  {NoStop}%
\bibitem [{\citenamefont {Allada}\ \emph {et~al.}(2014)\citenamefont {Allada}
  \emph {et~al.}}]{JeffersonLabHallA:2013mjr}%
  \BibitemOpen
  \bibfield  {author} {\bibinfo {author} {\bibfnamefont {K.}~\bibnamefont
  {Allada}} \emph {et~al.} (\bibinfo {collaboration} {Jefferson Lab Hall A}),\
  }\href {\doibase 10.1103/PhysRevC.89.042201} {\bibfield  {journal} {\bibinfo
  {journal} {Phys. Rev. C}\ }\textbf {\bibinfo {volume} {89}},\ \bibinfo
  {pages} {042201} (\bibinfo {year} {2014})},\ \Eprint
  {http://arxiv.org/abs/1311.1866} {arXiv:1311.1866 [nucl-ex]} \BibitemShut
  {NoStop}%
\bibitem [{\citenamefont {Alekseev}\ \emph {et~al.}(2010)\citenamefont
  {Alekseev} \emph {et~al.}}]{COMPASS:2010hbb}%
  \BibitemOpen
  \bibfield  {author} {\bibinfo {author} {\bibfnamefont {M.~G.}\ \bibnamefont
  {Alekseev}} \emph {et~al.} (\bibinfo {collaboration} {COMPASS}),\ }\href
  {\doibase 10.1016/j.physletb.2010.08.001} {\bibfield  {journal} {\bibinfo
  {journal} {Phys. Lett. B}\ }\textbf {\bibinfo {volume} {692}},\ \bibinfo
  {pages} {240} (\bibinfo {year} {2010})},\ \Eprint
  {http://arxiv.org/abs/1005.5609} {arXiv:1005.5609 [hep-ex]} \BibitemShut
  {NoStop}%
\bibitem [{\citenamefont {Accardi}\ \emph {et~al.}(2016)\citenamefont {Accardi}
  \emph {et~al.}}]{Accardi:2012qut}%
  \BibitemOpen
  \bibfield  {author} {\bibinfo {author} {\bibfnamefont {A.}~\bibnamefont
  {Accardi}} \emph {et~al.},\ }\href {\doibase 10.1140/epja/i2016-16268-9}
  {\bibfield  {journal} {\bibinfo  {journal} {Eur. Phys. J. A}\ }\textbf
  {\bibinfo {volume} {52}},\ \bibinfo {pages} {268} (\bibinfo {year} {2016})},\
  \Eprint {http://arxiv.org/abs/1212.1701} {arXiv:1212.1701 [nucl-ex]}
  \BibitemShut {NoStop}%
\bibitem [{\citenamefont {Abdul~Khalek}\ \emph {et~al.}(2021)\citenamefont
  {Abdul~Khalek} \emph {et~al.}}]{AbdulKhalek:2021gbh}%
  \BibitemOpen
  \bibfield  {author} {\bibinfo {author} {\bibfnamefont {R.}~\bibnamefont
  {Abdul~Khalek}} \emph {et~al.},\ }\href@noop {} {\  (\bibinfo {year}
  {2021})},\ \Eprint {http://arxiv.org/abs/2103.05419} {arXiv:2103.05419
  [physics.ins-det]} \BibitemShut {NoStop}%
\bibitem [{\citenamefont {Barone}\ \emph
  {et~al.}(2010{\natexlab{a}})\citenamefont {Barone}, \citenamefont
  {Bradamante},\ and\ \citenamefont {Martin}}]{Barone:2010zz}%
  \BibitemOpen
  \bibfield  {author} {\bibinfo {author} {\bibfnamefont {V.}~\bibnamefont
  {Barone}}, \bibinfo {author} {\bibfnamefont {F.}~\bibnamefont {Bradamante}},
  \ and\ \bibinfo {author} {\bibfnamefont {A.}~\bibnamefont {Martin}},\ }\href
  {\doibase 10.1016/j.ppnp.2010.07.003} {\bibfield  {journal} {\bibinfo
  {journal} {Prog. Part. Nucl. Phys.}\ }\textbf {\bibinfo {volume} {65}},\
  \bibinfo {pages} {267} (\bibinfo {year} {2010}{\natexlab{a}})},\ \Eprint
  {http://arxiv.org/abs/1011.0909} {arXiv:1011.0909 [hep-ph]} \BibitemShut
  {NoStop}%
\bibitem [{\citenamefont {Brodsky}\ \emph {et~al.}(2002)\citenamefont
  {Brodsky}, \citenamefont {Hwang},\ and\ \citenamefont
  {Schmidt}}]{Brodsky:2002cx}%
  \BibitemOpen
  \bibfield  {author} {\bibinfo {author} {\bibfnamefont {S.~J.}\ \bibnamefont
  {Brodsky}}, \bibinfo {author} {\bibfnamefont {D.~S.}\ \bibnamefont {Hwang}},
  \ and\ \bibinfo {author} {\bibfnamefont {I.}~\bibnamefont {Schmidt}},\ }\href
  {\doibase 10.1016/S0370-2693(02)01320-5} {\bibfield  {journal} {\bibinfo
  {journal} {Phys. Lett. B}\ }\textbf {\bibinfo {volume} {530}},\ \bibinfo
  {pages} {99} (\bibinfo {year} {2002})},\ \Eprint
  {http://arxiv.org/abs/hep-ph/0201296} {arXiv:hep-ph/0201296} \BibitemShut
  {NoStop}%
\bibitem [{\citenamefont {Buffing}\ \emph {et~al.}(2013)\citenamefont
  {Buffing}, \citenamefont {Mukherjee},\ and\ \citenamefont
  {Mulders}}]{Buffing:2013kca}%
  \BibitemOpen
  \bibfield  {author} {\bibinfo {author} {\bibfnamefont {M.~G.~A.}\
  \bibnamefont {Buffing}}, \bibinfo {author} {\bibfnamefont {A.}~\bibnamefont
  {Mukherjee}}, \ and\ \bibinfo {author} {\bibfnamefont {P.~J.}\ \bibnamefont
  {Mulders}},\ }\href {\doibase 10.1103/PhysRevD.88.054027} {\bibfield
  {journal} {\bibinfo  {journal} {Phys. Rev. D}\ }\textbf {\bibinfo {volume}
  {88}},\ \bibinfo {pages} {054027} (\bibinfo {year} {2013})},\ \Eprint
  {http://arxiv.org/abs/1306.5897} {arXiv:1306.5897 [hep-ph]} \BibitemShut
  {NoStop}%
\bibitem [{\citenamefont {Collins}(2002)}]{Collins:2002kn}%
  \BibitemOpen
  \bibfield  {author} {\bibinfo {author} {\bibfnamefont {J.~C.}\ \bibnamefont
  {Collins}},\ }\href {\doibase 10.1016/S0370-2693(02)01819-1} {\bibfield
  {journal} {\bibinfo  {journal} {Phys. Lett. B}\ }\textbf {\bibinfo {volume}
  {536}},\ \bibinfo {pages} {43} (\bibinfo {year} {2002})},\ \Eprint
  {http://arxiv.org/abs/hep-ph/0204004} {arXiv:hep-ph/0204004} \BibitemShut
  {NoStop}%
\bibitem [{\citenamefont {D'Alesio}\ \emph {et~al.}(2015)\citenamefont
  {D'Alesio}, \citenamefont {Murgia},\ and\ \citenamefont
  {Pisano}}]{DAlesio:2015fwo}%
  \BibitemOpen
  \bibfield  {author} {\bibinfo {author} {\bibfnamefont {U.}~\bibnamefont
  {D'Alesio}}, \bibinfo {author} {\bibfnamefont {F.}~\bibnamefont {Murgia}}, \
  and\ \bibinfo {author} {\bibfnamefont {C.}~\bibnamefont {Pisano}},\ }\href
  {\doibase 10.1007/JHEP09(2015)119} {\bibfield  {journal} {\bibinfo  {journal}
  {JHEP}\ }\textbf {\bibinfo {volume} {09}},\ \bibinfo {pages} {119} (\bibinfo
  {year} {2015})},\ \Eprint {http://arxiv.org/abs/1506.03078} {arXiv:1506.03078
  [hep-ph]} \BibitemShut {NoStop}%
\bibitem [{\citenamefont {Adolph}\ \emph {et~al.}(2017)\citenamefont {Adolph}
  \emph {et~al.}}]{COMPASS:2017ezz}%
  \BibitemOpen
  \bibfield  {author} {\bibinfo {author} {\bibfnamefont {C.}~\bibnamefont
  {Adolph}} \emph {et~al.} (\bibinfo {collaboration} {COMPASS}),\ }\href
  {\doibase 10.1016/j.physletb.2017.07.018} {\bibfield  {journal} {\bibinfo
  {journal} {Phys. Lett. B}\ }\textbf {\bibinfo {volume} {772}},\ \bibinfo
  {pages} {854} (\bibinfo {year} {2017})},\ \Eprint
  {http://arxiv.org/abs/1701.02453} {arXiv:1701.02453 [hep-ex]} \BibitemShut
  {NoStop}%
\bibitem [{\citenamefont {Sivers}(2006)}]{Sivers:2006rg}%
  \BibitemOpen
  \bibfield  {author} {\bibinfo {author} {\bibfnamefont {D.}~\bibnamefont
  {Sivers}},\ }\href {\doibase 10.1103/PhysRevD.74.094008} {\bibfield
  {journal} {\bibinfo  {journal} {Phys. Rev. D}\ }\textbf {\bibinfo {volume}
  {74}},\ \bibinfo {pages} {094008} (\bibinfo {year} {2006})},\ \Eprint
  {http://arxiv.org/abs/hep-ph/0609080} {arXiv:hep-ph/0609080} \BibitemShut
  {NoStop}%
\bibitem [{\citenamefont {Bacchetta}\ and\ \citenamefont
  {Radici}(2011)}]{Bacchetta:2011gx}%
  \BibitemOpen
  \bibfield  {author} {\bibinfo {author} {\bibfnamefont {A.}~\bibnamefont
  {Bacchetta}}\ and\ \bibinfo {author} {\bibfnamefont {M.}~\bibnamefont
  {Radici}},\ }\href {\doibase 10.1103/PhysRevLett.107.212001} {\bibfield
  {journal} {\bibinfo  {journal} {Phys. Rev. Lett.}\ }\textbf {\bibinfo
  {volume} {107}},\ \bibinfo {pages} {212001} (\bibinfo {year} {2011})},\
  \Eprint {http://arxiv.org/abs/1107.5755} {arXiv:1107.5755 [hep-ph]}
  \BibitemShut {NoStop}%
\bibitem [{\citenamefont {Anselmino}\ \emph {et~al.}(2005)\citenamefont
  {Anselmino}, \citenamefont {Boglione}, \citenamefont {D'Alesio},
  \citenamefont {Kotzinian}, \citenamefont {Murgia},\ and\ \citenamefont
  {Prokudin}}]{Anselmino:2005ea}%
  \BibitemOpen
  \bibfield  {author} {\bibinfo {author} {\bibfnamefont {M.}~\bibnamefont
  {Anselmino}}, \bibinfo {author} {\bibfnamefont {M.}~\bibnamefont {Boglione}},
  \bibinfo {author} {\bibfnamefont {U.}~\bibnamefont {D'Alesio}}, \bibinfo
  {author} {\bibfnamefont {A.}~\bibnamefont {Kotzinian}}, \bibinfo {author}
  {\bibfnamefont {F.}~\bibnamefont {Murgia}}, \ and\ \bibinfo {author}
  {\bibfnamefont {A.}~\bibnamefont {Prokudin}},\ }\href {\doibase
  10.1103/PhysRevD.72.094007} {\bibfield  {journal} {\bibinfo  {journal} {Phys.
  Rev. D}\ }\textbf {\bibinfo {volume} {72}},\ \bibinfo {pages} {094007}
  (\bibinfo {year} {2005})},\ \bibinfo {note} {[Erratum: Phys.Rev.D 72, 099903
  (2005)]},\ \Eprint {http://arxiv.org/abs/hep-ph/0507181}
  {arXiv:hep-ph/0507181} \BibitemShut {NoStop}%
\bibitem [{\citenamefont {Anselmino}\ \emph {et~al.}(2009)\citenamefont
  {Anselmino}, \citenamefont {Boglione}, \citenamefont {D'Alesio},
  \citenamefont {Kotzinian}, \citenamefont {Melis}, \citenamefont {Murgia},
  \citenamefont {Prokudin},\ and\ \citenamefont {Turk}}]{Anselmino:2008sga}%
  \BibitemOpen
  \bibfield  {author} {\bibinfo {author} {\bibfnamefont {M.}~\bibnamefont
  {Anselmino}}, \bibinfo {author} {\bibfnamefont {M.}~\bibnamefont {Boglione}},
  \bibinfo {author} {\bibfnamefont {U.}~\bibnamefont {D'Alesio}}, \bibinfo
  {author} {\bibfnamefont {A.}~\bibnamefont {Kotzinian}}, \bibinfo {author}
  {\bibfnamefont {S.}~\bibnamefont {Melis}}, \bibinfo {author} {\bibfnamefont
  {F.}~\bibnamefont {Murgia}}, \bibinfo {author} {\bibfnamefont
  {A.}~\bibnamefont {Prokudin}}, \ and\ \bibinfo {author} {\bibfnamefont
  {C.}~\bibnamefont {Turk}},\ }\href {\doibase 10.1140/epja/i2008-10697-y}
  {\bibfield  {journal} {\bibinfo  {journal} {Eur. Phys. J. A}\ }\textbf
  {\bibinfo {volume} {39}},\ \bibinfo {pages} {89} (\bibinfo {year} {2009})},\
  \Eprint {http://arxiv.org/abs/0805.2677} {arXiv:0805.2677 [hep-ph]}
  \BibitemShut {NoStop}%
\bibitem [{\citenamefont {Barone}\ \emph
  {et~al.}(2010{\natexlab{b}})\citenamefont {Barone}, \citenamefont {Melis},\
  and\ \citenamefont {Prokudin}}]{Barone:2010gk}%
  \BibitemOpen
  \bibfield  {author} {\bibinfo {author} {\bibfnamefont {V.}~\bibnamefont
  {Barone}}, \bibinfo {author} {\bibfnamefont {S.}~\bibnamefont {Melis}}, \
  and\ \bibinfo {author} {\bibfnamefont {A.}~\bibnamefont {Prokudin}},\ }\href
  {\doibase 10.1103/PhysRevD.82.114025} {\bibfield  {journal} {\bibinfo
  {journal} {Phys. Rev. D}\ }\textbf {\bibinfo {volume} {82}},\ \bibinfo
  {pages} {114025} (\bibinfo {year} {2010}{\natexlab{b}})},\ \Eprint
  {http://arxiv.org/abs/1009.3423} {arXiv:1009.3423 [hep-ph]} \BibitemShut
  {NoStop}%
\bibitem [{\citenamefont {Echevarria}\ \emph {et~al.}(2021)\citenamefont
  {Echevarria}, \citenamefont {Kang},\ and\ \citenamefont
  {Terry}}]{Echevarria:2020hpy}%
  \BibitemOpen
  \bibfield  {author} {\bibinfo {author} {\bibfnamefont {M.~G.}\ \bibnamefont
  {Echevarria}}, \bibinfo {author} {\bibfnamefont {Z.-B.}\ \bibnamefont
  {Kang}}, \ and\ \bibinfo {author} {\bibfnamefont {J.}~\bibnamefont {Terry}},\
  }\href {\doibase 10.1007/JHEP01(2021)126} {\bibfield  {journal} {\bibinfo
  {journal} {JHEP}\ }\textbf {\bibinfo {volume} {01}},\ \bibinfo {pages} {126}
  (\bibinfo {year} {2021})},\ \Eprint {http://arxiv.org/abs/2009.10710}
  {arXiv:2009.10710 [hep-ph]} \BibitemShut {NoStop}%
\bibitem [{\citenamefont {Bury}\ \emph {et~al.}(2021)\citenamefont {Bury},
  \citenamefont {Prokudin},\ and\ \citenamefont {Vladimirov}}]{Bury:2021sue}%
  \BibitemOpen
  \bibfield  {author} {\bibinfo {author} {\bibfnamefont {M.}~\bibnamefont
  {Bury}}, \bibinfo {author} {\bibfnamefont {A.}~\bibnamefont {Prokudin}}, \
  and\ \bibinfo {author} {\bibfnamefont {A.}~\bibnamefont {Vladimirov}},\
  }\href {\doibase 10.1007/JHEP05(2021)151} {\bibfield  {journal} {\bibinfo
  {journal} {JHEP}\ }\textbf {\bibinfo {volume} {05}},\ \bibinfo {pages} {151}
  (\bibinfo {year} {2021})},\ \Eprint {http://arxiv.org/abs/2103.03270}
  {arXiv:2103.03270 [hep-ph]} \BibitemShut {NoStop}%
\bibitem [{\citenamefont {Bacchetta}\ \emph {et~al.}(2022)\citenamefont
  {Bacchetta}, \citenamefont {Delcarro}, \citenamefont {Pisano},\ and\
  \citenamefont {Radici}}]{Bacchetta:2020gko}%
  \BibitemOpen
  \bibfield  {author} {\bibinfo {author} {\bibfnamefont {A.}~\bibnamefont
  {Bacchetta}}, \bibinfo {author} {\bibfnamefont {F.}~\bibnamefont {Delcarro}},
  \bibinfo {author} {\bibfnamefont {C.}~\bibnamefont {Pisano}}, \ and\ \bibinfo
  {author} {\bibfnamefont {M.}~\bibnamefont {Radici}},\ }\href {\doibase
  10.1016/j.physletb.2022.136961} {\bibfield  {journal} {\bibinfo  {journal}
  {Phys. Lett. B}\ }\textbf {\bibinfo {volume} {827}},\ \bibinfo {pages}
  {136961} (\bibinfo {year} {2022})},\ \Eprint
  {http://arxiv.org/abs/2004.14278} {arXiv:2004.14278 [hep-ph]} \BibitemShut
  {NoStop}%
\bibitem [{\citenamefont {Cammarota}\ \emph {et~al.}(2020)\citenamefont
  {Cammarota}, \citenamefont {Gamberg}, \citenamefont {Kang}, \citenamefont
  {Miller}, \citenamefont {Pitonyak}, \citenamefont {Prokudin}, \citenamefont
  {Rogers},\ and\ \citenamefont {Sato}}]{Cammarota:2020qcw}%
  \BibitemOpen
  \bibfield  {author} {\bibinfo {author} {\bibfnamefont {J.}~\bibnamefont
  {Cammarota}}, \bibinfo {author} {\bibfnamefont {L.}~\bibnamefont {Gamberg}},
  \bibinfo {author} {\bibfnamefont {Z.-B.}\ \bibnamefont {Kang}}, \bibinfo
  {author} {\bibfnamefont {J.~A.}\ \bibnamefont {Miller}}, \bibinfo {author}
  {\bibfnamefont {D.}~\bibnamefont {Pitonyak}}, \bibinfo {author}
  {\bibfnamefont {A.}~\bibnamefont {Prokudin}}, \bibinfo {author}
  {\bibfnamefont {T.~C.}\ \bibnamefont {Rogers}}, \ and\ \bibinfo {author}
  {\bibfnamefont {N.}~\bibnamefont {Sato}} (\bibinfo {collaboration} {Jefferson
  Lab Angular Momentum}),\ }\href {\doibase 10.1103/PhysRevD.102.054002}
  {\bibfield  {journal} {\bibinfo  {journal} {Phys. Rev. D}\ }\textbf {\bibinfo
  {volume} {102}},\ \bibinfo {pages} {054002} (\bibinfo {year} {2020})},\
  \Eprint {http://arxiv.org/abs/2002.08384} {arXiv:2002.08384 [hep-ph]}
  \BibitemShut {NoStop}%
\bibitem [{\citenamefont {Gamberg}\ \emph {et~al.}(2022)\citenamefont
  {Gamberg}, \citenamefont {Malda}, \citenamefont {Miller}, \citenamefont
  {Pitonyak}, \citenamefont {Prokudin},\ and\ \citenamefont
  {Sato}}]{Gamberg:2022kdb}%
  \BibitemOpen
  \bibfield  {author} {\bibinfo {author} {\bibfnamefont {L.}~\bibnamefont
  {Gamberg}}, \bibinfo {author} {\bibfnamefont {M.}~\bibnamefont {Malda}},
  \bibinfo {author} {\bibfnamefont {J.~A.}\ \bibnamefont {Miller}}, \bibinfo
  {author} {\bibfnamefont {D.}~\bibnamefont {Pitonyak}}, \bibinfo {author}
  {\bibfnamefont {A.}~\bibnamefont {Prokudin}}, \ and\ \bibinfo {author}
  {\bibfnamefont {N.}~\bibnamefont {Sato}} (\bibinfo {collaboration} {Jefferson
  Lab Angular Momentum (JAM), Jefferson Lab Angular Momentum}),\ }\href
  {\doibase 10.1103/PhysRevD.106.034014} {\bibfield  {journal} {\bibinfo
  {journal} {Phys. Rev. D}\ }\textbf {\bibinfo {volume} {106}},\ \bibinfo
  {pages} {034014} (\bibinfo {year} {2022})},\ \Eprint
  {http://arxiv.org/abs/2205.00999} {arXiv:2205.00999 [hep-ph]} \BibitemShut
  {NoStop}%
\bibitem [{\citenamefont {Boer}\ \emph {et~al.}(2015)\citenamefont {Boer},
  \citenamefont {Lorc\'e}, \citenamefont {Pisano},\ and\ \citenamefont
  {Zhou}}]{Boer:2015vso}%
  \BibitemOpen
  \bibfield  {author} {\bibinfo {author} {\bibfnamefont {D.}~\bibnamefont
  {Boer}}, \bibinfo {author} {\bibfnamefont {C.}~\bibnamefont {Lorc\'e}},
  \bibinfo {author} {\bibfnamefont {C.}~\bibnamefont {Pisano}}, \ and\ \bibinfo
  {author} {\bibfnamefont {J.}~\bibnamefont {Zhou}},\ }\href {\doibase
  10.1155/2015/371396} {\bibfield  {journal} {\bibinfo  {journal} {Adv. High
  Energy Phys.}\ }\textbf {\bibinfo {volume} {2015}},\ \bibinfo {pages}
  {371396} (\bibinfo {year} {2015})},\ \Eprint
  {http://arxiv.org/abs/1504.04332} {arXiv:1504.04332 [hep-ph]} \BibitemShut
  {NoStop}%
\bibitem [{\citenamefont {D'Alesio}\ \emph {et~al.}(2019)\citenamefont
  {D'Alesio}, \citenamefont {Flore}, \citenamefont {Murgia}, \citenamefont
  {Pisano},\ and\ \citenamefont {Taels}}]{DAlesio:2018rnv}%
  \BibitemOpen
  \bibfield  {author} {\bibinfo {author} {\bibfnamefont {U.}~\bibnamefont
  {D'Alesio}}, \bibinfo {author} {\bibfnamefont {C.}~\bibnamefont {Flore}},
  \bibinfo {author} {\bibfnamefont {F.}~\bibnamefont {Murgia}}, \bibinfo
  {author} {\bibfnamefont {C.}~\bibnamefont {Pisano}}, \ and\ \bibinfo {author}
  {\bibfnamefont {P.}~\bibnamefont {Taels}},\ }\href {\doibase
  10.1103/PhysRevD.99.036013} {\bibfield  {journal} {\bibinfo  {journal} {Phys.
  Rev. D}\ }\textbf {\bibinfo {volume} {99}},\ \bibinfo {pages} {036013}
  (\bibinfo {year} {2019})},\ \Eprint {http://arxiv.org/abs/1811.02970}
  {arXiv:1811.02970 [hep-ph]} \BibitemShut {NoStop}%
\bibitem [{\citenamefont {Airapetian}\ \emph {et~al.}(2009)\citenamefont
  {Airapetian} \emph {et~al.}}]{HERMES:2009lmz}%
  \BibitemOpen
  \bibfield  {author} {\bibinfo {author} {\bibfnamefont {A.}~\bibnamefont
  {Airapetian}} \emph {et~al.} (\bibinfo {collaboration} {HERMES}),\ }\href
  {\doibase 10.1103/PhysRevLett.103.152002} {\bibfield  {journal} {\bibinfo
  {journal} {Phys. Rev. Lett.}\ }\textbf {\bibinfo {volume} {103}},\ \bibinfo
  {pages} {152002} (\bibinfo {year} {2009})},\ \Eprint
  {http://arxiv.org/abs/0906.3918} {arXiv:0906.3918 [hep-ex]} \BibitemShut
  {NoStop}%
\bibitem [{\citenamefont {Alexakhin}\ \emph {et~al.}(2005)\citenamefont
  {Alexakhin} \emph {et~al.}}]{COMPASS:2005csq}%
  \BibitemOpen
  \bibfield  {author} {\bibinfo {author} {\bibfnamefont {V.~Y.}\ \bibnamefont
  {Alexakhin}} \emph {et~al.} (\bibinfo {collaboration} {COMPASS}),\ }\href
  {\doibase 10.1103/PhysRevLett.94.202002} {\bibfield  {journal} {\bibinfo
  {journal} {Phys. Rev. Lett.}\ }\textbf {\bibinfo {volume} {94}},\ \bibinfo
  {pages} {202002} (\bibinfo {year} {2005})},\ \Eprint
  {http://arxiv.org/abs/hep-ex/0503002} {arXiv:hep-ex/0503002} \BibitemShut
  {NoStop}%
\bibitem [{\citenamefont {Brodsky}\ and\ \citenamefont
  {Gardner}(2006)}]{Brodsky:2006ha}%
  \BibitemOpen
  \bibfield  {author} {\bibinfo {author} {\bibfnamefont {S.~J.}\ \bibnamefont
  {Brodsky}}\ and\ \bibinfo {author} {\bibfnamefont {S.}~\bibnamefont
  {Gardner}},\ }\href {\doibase 10.1016/j.physletb.2006.10.024} {\bibfield
  {journal} {\bibinfo  {journal} {Phys. Lett. B}\ }\textbf {\bibinfo {volume}
  {643}},\ \bibinfo {pages} {22} (\bibinfo {year} {2006})},\ \Eprint
  {http://arxiv.org/abs/hep-ph/0608219} {arXiv:hep-ph/0608219} \BibitemShut
  {NoStop}%
\bibitem [{\citenamefont {Burkardt}(2004)}]{Burkardt:2004ur}%
  \BibitemOpen
  \bibfield  {author} {\bibinfo {author} {\bibfnamefont {M.}~\bibnamefont
  {Burkardt}},\ }\href {\doibase 10.1103/PhysRevD.69.091501} {\bibfield
  {journal} {\bibinfo  {journal} {Phys. Rev. D}\ }\textbf {\bibinfo {volume}
  {69}},\ \bibinfo {pages} {091501} (\bibinfo {year} {2004})},\ \Eprint
  {http://arxiv.org/abs/hep-ph/0402014} {arXiv:hep-ph/0402014} \BibitemShut
  {NoStop}%
\bibitem [{\citenamefont {Aidala}\ \emph {et~al.}(2017)\citenamefont {Aidala}
  \emph {et~al.}}]{PHENIX:2017wbv}%
  \BibitemOpen
  \bibfield  {author} {\bibinfo {author} {\bibfnamefont {C.}~\bibnamefont
  {Aidala}} \emph {et~al.} (\bibinfo {collaboration} {PHENIX}),\ }\href
  {\doibase 10.1103/PhysRevD.95.112001} {\bibfield  {journal} {\bibinfo
  {journal} {Phys. Rev. D}\ }\textbf {\bibinfo {volume} {95}},\ \bibinfo
  {pages} {112001} (\bibinfo {year} {2017})},\ \Eprint
  {http://arxiv.org/abs/1703.09333} {arXiv:1703.09333 [hep-ex]} \BibitemShut
  {NoStop}%
\bibitem [{\citenamefont {Mulders}\ and\ \citenamefont
  {Rodrigues}(2001)}]{Mulders:2000sh}%
  \BibitemOpen
  \bibfield  {author} {\bibinfo {author} {\bibfnamefont {P.~J.}\ \bibnamefont
  {Mulders}}\ and\ \bibinfo {author} {\bibfnamefont {J.}~\bibnamefont
  {Rodrigues}},\ }\href {\doibase 10.1103/PhysRevD.63.094021} {\bibfield
  {journal} {\bibinfo  {journal} {Phys. Rev. D}\ }\textbf {\bibinfo {volume}
  {63}},\ \bibinfo {pages} {094021} (\bibinfo {year} {2001})},\ \Eprint
  {http://arxiv.org/abs/hep-ph/0009343} {arXiv:hep-ph/0009343} \BibitemShut
  {NoStop}%
\bibitem [{\citenamefont {Zheng}\ \emph {et~al.}(2018)\citenamefont {Zheng},
  \citenamefont {Aschenauer}, \citenamefont {Lee}, \citenamefont {Xiao},\ and\
  \citenamefont {Yin}}]{Zheng:2018ssm}%
  \BibitemOpen
  \bibfield  {author} {\bibinfo {author} {\bibfnamefont {L.}~\bibnamefont
  {Zheng}}, \bibinfo {author} {\bibfnamefont {E.~C.}\ \bibnamefont
  {Aschenauer}}, \bibinfo {author} {\bibfnamefont {J.~H.}\ \bibnamefont {Lee}},
  \bibinfo {author} {\bibfnamefont {B.-W.}\ \bibnamefont {Xiao}}, \ and\
  \bibinfo {author} {\bibfnamefont {Z.-B.}\ \bibnamefont {Yin}},\ }\href
  {\doibase 10.1103/PhysRevD.98.034011} {\bibfield  {journal} {\bibinfo
  {journal} {Phys. Rev. D}\ }\textbf {\bibinfo {volume} {98}},\ \bibinfo
  {pages} {034011} (\bibinfo {year} {2018})},\ \Eprint
  {http://arxiv.org/abs/1805.05290} {arXiv:1805.05290 [hep-ph]} \BibitemShut
  {NoStop}%
\bibitem [{\citenamefont {Ji}\ \emph {et~al.}(2005)\citenamefont {Ji},
  \citenamefont {Ma},\ and\ \citenamefont {Yuan}}]{Ji:2004wu}%
  \BibitemOpen
  \bibfield  {author} {\bibinfo {author} {\bibfnamefont {X.-d.}\ \bibnamefont
  {Ji}}, \bibinfo {author} {\bibfnamefont {J.-p.}\ \bibnamefont {Ma}}, \ and\
  \bibinfo {author} {\bibfnamefont {F.}~\bibnamefont {Yuan}},\ }\href {\doibase
  10.1103/PhysRevD.71.034005} {\bibfield  {journal} {\bibinfo  {journal} {Phys.
  Rev. D}\ }\textbf {\bibinfo {volume} {71}},\ \bibinfo {pages} {034005}
  (\bibinfo {year} {2005})},\ \Eprint {http://arxiv.org/abs/hep-ph/0404183}
  {arXiv:hep-ph/0404183} \BibitemShut {NoStop}%
\bibitem [{\citenamefont {Boer}\ \emph {et~al.}(2016)\citenamefont {Boer},
  \citenamefont {Mulders}, \citenamefont {Pisano},\ and\ \citenamefont
  {Zhou}}]{Boer:2016fqd}%
  \BibitemOpen
  \bibfield  {author} {\bibinfo {author} {\bibfnamefont {D.}~\bibnamefont
  {Boer}}, \bibinfo {author} {\bibfnamefont {P.~J.}\ \bibnamefont {Mulders}},
  \bibinfo {author} {\bibfnamefont {C.}~\bibnamefont {Pisano}}, \ and\ \bibinfo
  {author} {\bibfnamefont {J.}~\bibnamefont {Zhou}},\ }\href {\doibase
  10.1007/JHEP08(2016)001} {\bibfield  {journal} {\bibinfo  {journal} {JHEP}\
  }\textbf {\bibinfo {volume} {08}},\ \bibinfo {pages} {001} (\bibinfo {year}
  {2016})},\ \Eprint {http://arxiv.org/abs/1605.07934} {arXiv:1605.07934
  [hep-ph]} \BibitemShut {NoStop}%
\bibitem [{\citenamefont {Vogt}(2018)}]{Vogt:2018oje}%
  \BibitemOpen
  \bibfield  {author} {\bibinfo {author} {\bibfnamefont {R.}~\bibnamefont
  {Vogt}},\ }\href {\doibase 10.1103/PhysRevC.98.034907} {\bibfield  {journal}
  {\bibinfo  {journal} {Phys. Rev. C}\ }\textbf {\bibinfo {volume} {98}},\
  \bibinfo {pages} {034907} (\bibinfo {year} {2018})},\ \Eprint
  {http://arxiv.org/abs/1806.01904} {arXiv:1806.01904 [hep-ph]} \BibitemShut
  {NoStop}%
\bibitem [{\citenamefont {Citron}\ \emph {et~al.}(2019)\citenamefont {Citron}
  \emph {et~al.}}]{Citron:2018lsq}%
  \BibitemOpen
  \bibfield  {author} {\bibinfo {author} {\bibfnamefont {Z.}~\bibnamefont
  {Citron}} \emph {et~al.},\ }\href {\doibase 10.23731/CYRM-2019-007.1159}
  {\bibfield  {journal} {\bibinfo  {journal} {CERN Yellow Rep. Monogr.}\
  }\textbf {\bibinfo {volume} {7}},\ \bibinfo {pages} {1159} (\bibinfo {year}
  {2019})},\ \Eprint {http://arxiv.org/abs/1812.06772} {arXiv:1812.06772
  [hep-ph]} \BibitemShut {NoStop}%
\bibitem [{\citenamefont {Arrington}\ \emph {et~al.}(2021)\citenamefont
  {Arrington} \emph {et~al.}}]{Arrington:2021yeb}%
  \BibitemOpen
  \bibfield  {author} {\bibinfo {author} {\bibfnamefont {J.}~\bibnamefont
  {Arrington}} \emph {et~al.},\ }\href@noop {} {\  (\bibinfo {year} {2021})},\
  \Eprint {http://arxiv.org/abs/2102.08337} {arXiv:2102.08337 [nucl-ex]}
  \BibitemShut {NoStop}%
\bibitem [{\citenamefont {Contin}\ \emph {et~al.}(2018)\citenamefont {Contin}
  \emph {et~al.}}]{Contin:2017mck}%
  \BibitemOpen
  \bibfield  {author} {\bibinfo {author} {\bibfnamefont {G.}~\bibnamefont
  {Contin}} \emph {et~al.},\ }\href {\doibase 10.1016/j.nima.2018.03.003}
  {\bibfield  {journal} {\bibinfo  {journal} {Nucl. Instrum. Meth. A}\ }\textbf
  {\bibinfo {volume} {907}},\ \bibinfo {pages} {60} (\bibinfo {year} {2018})},\
  \Eprint {http://arxiv.org/abs/1710.02176} {arXiv:1710.02176
  [physics.ins-det]} \BibitemShut {NoStop}%
\bibitem [{\citenamefont {Fantoni}(2020)}]{Fantoni:2020iyr}%
  \BibitemOpen
  \bibfield  {author} {\bibinfo {author} {\bibfnamefont {A.}~\bibnamefont
  {Fantoni}} (\bibinfo {collaboration} {ALICE}),\ }\href {\doibase
  10.1088/1402-4896/aba0f7} {\bibfield  {journal} {\bibinfo  {journal} {Phys.
  Scripta}\ }\textbf {\bibinfo {volume} {95}},\ \bibinfo {pages} {084011}
  (\bibinfo {year} {2020})}\BibitemShut {NoStop}%
\bibitem [{\citenamefont {{ECCE Collaboration,
  https://www.ecce-eic.org}}()}]{Ecce:2021ec}%
  \BibitemOpen
  \bibfield  {author} {\bibinfo {author} {\bibnamefont {{ECCE Collaboration,
  https://www.ecce-eic.org}}},\ }\href@noop {} {\ }\BibitemShut {NoStop}%
\bibitem [{\citenamefont {{ATHENA Collaboration, \\
  https://wiki.bnl.gov/athena/index.php}}()}]{Athena:2021at}%
  \BibitemOpen
  \bibfield  {author} {\bibinfo {author} {\bibnamefont {{ATHENA Collaboration,
  \\ https://wiki.bnl.gov/athena/index.php}}},\ }\href@noop {} {\ }\BibitemShut
  {NoStop}%
\bibitem [{\citenamefont {Agostinelli}\ \emph {et~al.}(2003)\citenamefont
  {Agostinelli} \emph {et~al.}}]{GEANT4:2002zbu}%
  \BibitemOpen
  \bibfield  {author} {\bibinfo {author} {\bibfnamefont {S.}~\bibnamefont
  {Agostinelli}} \emph {et~al.} (\bibinfo {collaboration} {GEANT4}),\ }\href
  {\doibase 10.1016/S0168-9002(03)01368-8} {\bibfield  {journal} {\bibinfo
  {journal} {Nucl. Instrum. Meth. A}\ }\textbf {\bibinfo {volume} {506}},\
  \bibinfo {pages} {250} (\bibinfo {year} {2003})}\BibitemShut {NoStop}%
\bibitem [{\citenamefont {Kelsey}\ \emph {et~al.}(2021)\citenamefont {Kelsey},
  \citenamefont {Cruz-Torres}, \citenamefont {Dong}, \citenamefont {Ji},
  \citenamefont {Radhakrishnan},\ and\ \citenamefont
  {Sichtermann}}]{Kelsey:2021gpk}%
  \BibitemOpen
  \bibfield  {author} {\bibinfo {author} {\bibfnamefont {M.}~\bibnamefont
  {Kelsey}}, \bibinfo {author} {\bibfnamefont {R.}~\bibnamefont {Cruz-Torres}},
  \bibinfo {author} {\bibfnamefont {X.}~\bibnamefont {Dong}}, \bibinfo {author}
  {\bibfnamefont {Y.}~\bibnamefont {Ji}}, \bibinfo {author} {\bibfnamefont
  {S.}~\bibnamefont {Radhakrishnan}}, \ and\ \bibinfo {author} {\bibfnamefont
  {E.}~\bibnamefont {Sichtermann}},\ }\href {\doibase
  10.1103/PhysRevD.104.054002} {\bibfield  {journal} {\bibinfo  {journal}
  {Phys. Rev. D}\ }\textbf {\bibinfo {volume} {104}},\ \bibinfo {pages}
  {054002} (\bibinfo {year} {2021})},\ \Eprint
  {http://arxiv.org/abs/2107.05632} {arXiv:2107.05632 [hep-ph]} \BibitemShut
  {NoStop}%
\bibitem [{\citenamefont {Anderle}\ \emph {et~al.}(2021)\citenamefont
  {Anderle}, \citenamefont {Dong}, \citenamefont {Hekhorn}, \citenamefont
  {Kelsey}, \citenamefont {Radhakrishnan}, \citenamefont {Sichtermann},
  \citenamefont {Xia}, \citenamefont {Xing}, \citenamefont {Yuan},\ and\
  \citenamefont {Zhao}}]{Anderle:2021hpa}%
  \BibitemOpen
  \bibfield  {author} {\bibinfo {author} {\bibfnamefont {D.~P.}\ \bibnamefont
  {Anderle}}, \bibinfo {author} {\bibfnamefont {X.}~\bibnamefont {Dong}},
  \bibinfo {author} {\bibfnamefont {F.}~\bibnamefont {Hekhorn}}, \bibinfo
  {author} {\bibfnamefont {M.}~\bibnamefont {Kelsey}}, \bibinfo {author}
  {\bibfnamefont {S.}~\bibnamefont {Radhakrishnan}}, \bibinfo {author}
  {\bibfnamefont {E.}~\bibnamefont {Sichtermann}}, \bibinfo {author}
  {\bibfnamefont {L.}~\bibnamefont {Xia}}, \bibinfo {author} {\bibfnamefont
  {H.}~\bibnamefont {Xing}}, \bibinfo {author} {\bibfnamefont {F.}~\bibnamefont
  {Yuan}}, \ and\ \bibinfo {author} {\bibfnamefont {Y.}~\bibnamefont {Zhao}},\
  }\href {\doibase 10.1103/PhysRevD.104.114039} {\bibfield  {journal} {\bibinfo
   {journal} {Phys. Rev. D}\ }\textbf {\bibinfo {volume} {104}},\ \bibinfo
  {pages} {114039} (\bibinfo {year} {2021})},\ \Eprint
  {http://arxiv.org/abs/2110.04489} {arXiv:2110.04489 [hep-ex]} \BibitemShut
  {NoStop}%
\bibitem [{\citenamefont {Sjostrand}\ \emph {et~al.}(2006)\citenamefont
  {Sjostrand}, \citenamefont {Mrenna},\ and\ \citenamefont
  {Skands}}]{Sjostrand:2006za}%
  \BibitemOpen
  \bibfield  {author} {\bibinfo {author} {\bibfnamefont {T.}~\bibnamefont
  {Sjostrand}}, \bibinfo {author} {\bibfnamefont {S.}~\bibnamefont {Mrenna}}, \
  and\ \bibinfo {author} {\bibfnamefont {P.~Z.}\ \bibnamefont {Skands}},\
  }\href {\doibase 10.1088/1126-6708/2006/05/026} {\bibfield  {journal}
  {\bibinfo  {journal} {JHEP}\ }\textbf {\bibinfo {volume} {05}},\ \bibinfo
  {pages} {026} (\bibinfo {year} {2006})},\ \Eprint
  {http://arxiv.org/abs/hep-ph/0603175} {arXiv:hep-ph/0603175} \BibitemShut
  {NoStop}%
\bibitem [{\citenamefont {Zyla}\ \emph {et~al.}(2020)\citenamefont {Zyla} \emph
  {et~al.}}]{Zyla:2020zbs}%
  \BibitemOpen
  \bibfield  {author} {\bibinfo {author} {\bibfnamefont {P.}~\bibnamefont
  {Zyla}} \emph {et~al.} (\bibinfo {collaboration} {Particle Data Group}),\
  }\href {\doibase 10.1093/ptep/ptaa104} {\bibfield  {journal} {\bibinfo
  {journal} {PTEP}\ }\textbf {\bibinfo {volume} {2020}},\ \bibinfo {pages}
  {083C01} (\bibinfo {year} {2020})}\BibitemShut {NoStop}%
\bibitem [{\citenamefont {Cacciari}\ \emph {et~al.}(2008)\citenamefont
  {Cacciari}, \citenamefont {Salam},\ and\ \citenamefont
  {Soyez}}]{Cacciari:2008gp}%
  \BibitemOpen
  \bibfield  {author} {\bibinfo {author} {\bibfnamefont {M.}~\bibnamefont
  {Cacciari}}, \bibinfo {author} {\bibfnamefont {G.~P.}\ \bibnamefont {Salam}},
  \ and\ \bibinfo {author} {\bibfnamefont {G.}~\bibnamefont {Soyez}},\ }\href
  {\doibase 10.1088/1126-6708/2008/04/063} {\bibfield  {journal} {\bibinfo
  {journal} {JHEP}\ }\textbf {\bibinfo {volume} {04}},\ \bibinfo {pages} {063}
  (\bibinfo {year} {2008})},\ \Eprint {http://arxiv.org/abs/0802.1189}
  {arXiv:0802.1189 [hep-ph]} \BibitemShut {NoStop}%
\bibitem [{\citenamefont {Cacciari}\ \emph {et~al.}(2012)\citenamefont
  {Cacciari}, \citenamefont {Salam},\ and\ \citenamefont
  {Soyez}}]{Cacciari:2011ma}%
  \BibitemOpen
  \bibfield  {author} {\bibinfo {author} {\bibfnamefont {M.}~\bibnamefont
  {Cacciari}}, \bibinfo {author} {\bibfnamefont {G.~P.}\ \bibnamefont {Salam}},
  \ and\ \bibinfo {author} {\bibfnamefont {G.}~\bibnamefont {Soyez}},\ }\href
  {\doibase 10.1140/epjc/s10052-012-1896-2} {\bibfield  {journal} {\bibinfo
  {journal} {Eur. Phys. J. C}\ }\textbf {\bibinfo {volume} {72}},\ \bibinfo
  {pages} {1896} (\bibinfo {year} {2012})},\ \Eprint
  {http://arxiv.org/abs/1111.6097} {arXiv:1111.6097 [hep-ph]} \BibitemShut
  {NoStop}%
\bibitem [{\citenamefont {Hocker}\ \emph {et~al.}(2007)\citenamefont {Hocker}
  \emph {et~al.}}]{Hocker:2007ht}%
  \BibitemOpen
  \bibfield  {author} {\bibinfo {author} {\bibfnamefont {A.}~\bibnamefont
  {Hocker}} \emph {et~al.},\ }\href@noop {} {\  (\bibinfo {year} {2007})},\
  \Eprint {http://arxiv.org/abs/physics/0703039} {arXiv:physics/0703039}
  \BibitemShut {NoStop}%
\bibitem [{\citenamefont {Anselmino}\ \emph {et~al.}(2017)\citenamefont
  {Anselmino}, \citenamefont {Boglione}, \citenamefont {D'Alesio},
  \citenamefont {Murgia},\ and\ \citenamefont {Prokudin}}]{Anselmino:2016uie}%
  \BibitemOpen
  \bibfield  {author} {\bibinfo {author} {\bibfnamefont {M.}~\bibnamefont
  {Anselmino}}, \bibinfo {author} {\bibfnamefont {M.}~\bibnamefont {Boglione}},
  \bibinfo {author} {\bibfnamefont {U.}~\bibnamefont {D'Alesio}}, \bibinfo
  {author} {\bibfnamefont {F.}~\bibnamefont {Murgia}}, \ and\ \bibinfo {author}
  {\bibfnamefont {A.}~\bibnamefont {Prokudin}},\ }\href {\doibase
  10.1007/JHEP04(2017)046} {\bibfield  {journal} {\bibinfo  {journal} {JHEP}\
  }\textbf {\bibinfo {volume} {04}},\ \bibinfo {pages} {046} (\bibinfo {year}
  {2017})},\ \Eprint {http://arxiv.org/abs/1612.06413} {arXiv:1612.06413
  [hep-ph]} \BibitemShut {NoStop}%
\end{thebibliography}%

\end{document}